\documentclass[12pt]{article}
\pdfoutput=1
\usepackage{epsfig,amsfonts,amsthm}
\usepackage{amsmath,amssymb}
\usepackage{array}
\newcommand{\be}{\begin{equation}}
\newcommand{\ee}{\end{equation}}
\newcommand{\bea}{\begin{eqnarray}}
\newcommand{\eea}{\end{eqnarray}}

\renewcommand{\Re}{\mathrm{Re }}
\renewcommand{\Im}{\mathrm{Im }}
\newcommand{\doublet}[2]{ \left( \begin{array}{c}#1 \\ #2 \end{array}\right) }

\usepackage{color}

\def\lsim{\mathrel{\rlap{\lower4pt\hbox{\hskip1pt$\sim$}}
    \raise1pt\hbox{$<$}}}         
\def\gsim{\mathrel{\rlap{\lower4pt\hbox{\hskip1pt$\sim$}}
    \raise1pt\hbox{$>$}}}         

\def\beq{\begin{equation}}
\def\eeq{\end{equation}}
\def\bea{\begin{eqnarray}}
\def\eea{\end{eqnarray}}

\def\<{\left\langle}
\def\>{\right\rangle}

\newcommand{\bt}{\begin{tabular}}
\newcommand{\et}{\end{tabular}}

\allowdisplaybreaks[2]
\begin{document}
\bibliographystyle{OurBibTeX}

\title{\hfill ~\\[-30mm]
                  \textbf{CP Violating Two-Higgs-Doublet Model: Constraints and LHC Predictions
                }        }

\author{\\[-5mm]
Venus~Keus\footnote{E-mail: {\tt venus.keus@helsinki.fi}} $^{1,2}$,\ 
Stephen F. King\footnote{E-mail: {\tt king@soton.ac.uk}} $^{2}$,\ 
Stefano~Moretti\footnote{E-mail: {\tt s.moretti@soton.ac.uk}} $^{2,3}$ ,\
Kei~Yagyu\footnote{E-mail: {\tt k.yagyu@soton.ac.uk}} $^{2}$ 
\\ \\
  \emph{\small  $^1$  Department of Physics and Helsinki Institute of Physics,}\\
  \emph{\small Gustaf Hallstromin katu 2, FIN-00014 University of Helsinki, Finland}\\
  \emph{\small $^2$ School of Physics and Astronomy, University of Southampton,}\\
  \emph{\small Southampton, SO17 1BJ, United Kingdom}\\
  \emph{\small $^3$ Particle Physics Department, Rutherford Appleton Laboratory,}\\
  \emph{\small Chilton, Didcot, Oxon OX11 0QX, United Kingdom}\\[4mm]}

\maketitle

\begin{abstract}
\noindent
{Two-Higgs-Doublet Models (2HDMs) are amongst the simplest extensions of the Standard Model.
Such models allow for tree-level CP Violation (CPV) in the Higgs sector.
We analyse a class of CPV 2HDM (of Type-I) in which only one of the two Higgs doublets couples to 
quarks and leptons, avoiding dangerous Flavour Changing Neutral Currents. We provide an up to date and comprehensive analysis of the constraints and 
Large Hadron Collider (LHC) predictions of such a model. Of immediate interest to the LHC Run 2 is the golden channel where all three neutral
Higgs bosons are observed to decay into gauge boson pairs, $WW$ and $ZZ$, providing a smoking gun signature of the
CPV 2HDM.
} 
 \end{abstract}
\thispagestyle{empty}
\vfill
\newpage
\setcounter{page}{1}

\section{Introduction}

The Standard Model (SM) contains one Higgs doublet which is responsible for  Electro-Weak Symmetry Breaking (EWSB).
The corresponding Higgs boson, with a mass of $\approx 125$ GeV, was discovered in 2012 by the ATLAS and CMS experiments at the  Large Hadron Collider (LHC) \cite{Aad:2012tfa,Chatrchyan:2012ufa}. Although its properties agree so far with the predictions of the SM, 
including EW Precision Data (EWPD), 
it remains an intriguing possibility that the observed Higgs boson, denoted here as $h$, may just be one member of an  extended Higgs sector.
A good motivation for such an extended Higgs sector is the fact  that it allows for a new source of
CP Violation (CPV), 
as required to explain  the matter-antimatter asymmetry of the Universe.
Sakharov discovered that CPV is a necessary condition for  
matter-antimatter asymmetry generation \cite{Sakharov:1967dj} and it was later shown 
that CPV in the SM is insufficient for this purpose \cite{Kuzmin:1985mm}.

Among the simplest Higgs extensions are the Two-Higgs-Doublet Models (2HDMs), wherein the SM is extended with one extra Higgs doublet with the same quantum numbers as the SM one. 
CP Conserving (CPC) 2HDMs have been studied in detail in the literature \cite{Gunion:2002zf,Branco:2011iw}.
With the introduction of an extra Higgs doublet to which fermions can couple, one encounters the risk of introducing Flavour Changing Neutral Currents (FCNCs) at tree level, which are tightly constrained by experiment. However, these dangerous FCNCs can  be avoided by imposing a $Z_2$ symmetry on the scalar potential and assigning $Z_2$ charges to the fermions. Under this setup, there are four independent types of Yukawa interactions
which are the so-called Type-I, Type-II, Type-X and Type-Y\footnote{The Type-X and Type-Y 2HDMs 
are also referred to as the lepton-specific and flipped 2HDMs, respectively \cite{Branco:2011iw}.} \cite{four-types,typeX} depending on the $Z_2$ charge assignment to fermions. 

In a CPC 2HDM, one of the three states is identified as the CP-odd Higgs boson which does not couple to the gauge bosons. In a CPV 2HDM, however, all  three neutral Higgs states are mixed, one of which is identified with the 125 GeV Higgs bosons and all have non-zero Higgs-gauge-gauge type interactions.
One of the features of the CPV 2HDMs, then, is the mixing of the three neutral Higgs bosons. 
CPV 2HDMs have previously been studied in the literature 
(for early literature see \cite{Branco:2011iw}, \cite{Haber:2006ue} and references therein). 
Recently, in \cite{Sokolowska:2008bt,Grzadkowski:2014ada,Grzadkowski:2013rza,Grzadkowski:2015zma} model-independent approaches to CPV 2HDMs have been presented using the CP-odd weak-basis invariants.
Charged Higgs phenomenology in CPV 2HDMs has been considered in \cite{Jung:2010ik,Arhrib:2010ju,Bao:2010sz,Basso:2012st,Basso:2013wna,Mader:2012pm}.  
Surviving regions of the parameter space passing all experimental constraints in CPV 2HDMs have been studied in \cite{Jung:2013hka,Brod:2013cka,Inoue:2014nva,Gaitan:2015aia,Chen:2015gaa} and in \cite{Ipek:2013iba} with a focus on EW Baryogenesis.
Search signals for explicit CPV have been suggested for $Z_2$ symmetric 2HDMs in \cite{Barroso:2012wz,Fontes:2015xva} and for the general 2HDM in \cite{DiazCruz:2012xc}.

In the present paper, we provide a dedicated analysis of CPV in Type-I 2HDMs,
which updates and extends the discussions so far in the literature, including all the relevant constraints and LHC predictions.
We study explicit CPV in the case of a 2HDM with a softly-broken $Z_2$ symmetry where there is only one 
relevant complex parameter, namely $\lambda_5$\footnote{The imaginary part of the soft symmetry breaking term, $\mu_3^2$, can be written in terms of the imaginary part of $\lambda_5$.}. The imaginary part of $\lambda_5$ is constrained by Electric Dipole Moment (EDM) experiments, by EWPD, by unitarity and by vacuum stability constraints. We take into account all these constraints and parametrise CPV in the model in terms of the imaginary part of $\lambda_5$. 
We especially focus on the Type-I Yukawa interaction, where only one of the Higgs doublets couples to fermions and the extra Higgs boson couplings to fermions are suppressed by $1/\tan\beta$, 
where $\tan\beta$ is the ratio of two Vacuum Expectation Values (VEVs) of the two Higgs doublets. 
However, the extra Higgs bosons decays to $W^+W^-$ and $ZZ$ can be enhanced with large $\tan\beta$ due to suppressed decays to a fermion pair when the value of mixing angles and mass eigenvalues of the neutral Higgs states are fixed.
In other 2HDM types, some Yukawa couplings are proportional to $\tan\beta$ which leads to dominant fermion-pair decays of the neutral Higgses and could hide the $W^+W^-$ and $ZZ$ decay modes.
Moreover, in the Type-I 2HDM, extra Higgs boson contributions to EDMs are suppressed in the large $\tan\beta$ regime and mainly the modified couplings of the SM-like Higgs boson contribute to EDMs. 
We present LHC signatures for observing CPV in this model.
Of immediate interest to the LHC is the golden channel where all three neutral
Higgs bosons are observed to decay into weak gauge boson pairs, i.e., $W^+W^-$ and $ZZ$, providing a smoking gun signature of 
CPV 2HDMs (since purely CP-odd Higgs states cannot decay in these modes).
In summary, we perform a dedicated study of the CPV Type-I 2HDM  where we take into account the latest experimental and theoretical bounds and present the gauge couplings and Branching Ratios (BRs) of the neutral and charged Higgs bosons, the ratio of decay rates of the SM-like Higgs boson and Higgs signal strengths.

The remainder of this paper is organised as follows. 
In Section \ref{scalar} we present the scalar potential in $Z_2$-symmetric 2HDMs and the mass spectra in their CPC and CPV limits. 
In Section \ref{yukawa-kinetic} we show the Yukawa and kinetic Lagrangian in the CPV limit of the Type-I model. 
In Section \ref{constraints} 
we show the constraints imposed on the model and present four sets of parameters (mass spectra) 
allowed by these constraints for different values of $\tan\beta$ and $\sin(\beta-\tilde{\alpha})$ ($\tilde{\alpha}$ being
a mixing parameter). 
In the remainder of Section \ref{plots} we show the gauge couplings and Branching Ratios (BRs) of the neutral and charged Higgs bosons, 
the ratio of decay rates of the SM Higgs boson and  Higgs signal strengths in this model. We 
recap  our results and draw our conclusions in Section \ref{conclusion}.

\section{The scalar potential}
\label{scalar}

The most general 2HDM potential is of the following form:
\begin{align}
\label{potgen}
V^{gen} &= \mu^2_{1} (\phi_1^\dagger \phi_1) +\mu^2_2 (\phi_2^\dagger \phi_2) -\biggl[ \mu^2_3(\phi_1^\dagger \phi_2) + h.c. \biggr] \notag\\
&+ \frac{1}{2} \lambda_{1} (\phi_1^\dagger \phi_1)^2+\frac{1}{2} \lambda_{2} (\phi_2^\dagger \phi_2)^2  + \lambda_{3} (\phi_1^\dagger \phi_1)(\phi_2^\dagger \phi_2) 
 + \lambda_{4} (\phi_1^\dagger \phi_2)(\phi_2^\dagger \phi_1)\notag\\
&+  \biggl[\frac{1}{2} \lambda_{5}  (\phi_1^\dagger \phi_2)^2 + \lambda_{6} (\phi_1^\dagger \phi_1)(\phi_1^\dagger \phi_2)  
+ \lambda_{7} (\phi_2^\dagger \phi_2)(\phi_1^\dagger \phi_2) + h.c. \biggr]. 
\end{align}
In general, the scalar doublets are defined as
\be 
\phi_1= \doublet{$\begin{scriptsize}$ \phi^+_1 $\end{scriptsize}$}{\frac{v_1+h_1^0+ia^0_1}{\sqrt{2}}} ,\quad 
\phi_2= \doublet{$\begin{scriptsize}$ \phi^+_2 $\end{scriptsize}$}{\frac{v_2+h^0_2+ia^0_2}{\sqrt{2}}} ,
\label{fields}
\ee
where $v_1$ and $v_2$ could in principle be complex. 

In the general case, the 2HDMs suffer from the appearance of FCNCs at the tree level which are strongly restricted experimentally. 
It is known that imposing a $Z_2$ symmetry, which can be softly-broken in general,  
on the scalar potential and extending it to the fermion sector could forbid these FCNCs. 
Depending on the $Z_2$ charge assignment for fermions, 
four independent types of Yukawa interactions are allowed. We will discuss the types of Yukawa interactions in Section~\ref{yukawa-kinetic}. 
In the following, the transformations of two Higgs doublets under $Z_2$ are fixed to be $\phi_1\to +\phi_1$ and $\phi_2\to -\phi_2$. 

Imposing the softly-broken $Z_2$ symmetry 
on the potential reduces it to
\begin{align}
\label{pot1}
V &= \mu^2_{1} (\phi_1^\dagger \phi_1) +\mu^2_2 (\phi_2^\dagger \phi_2) -\biggl[ \mu^2_3(\phi_1^\dagger \phi_2) + h.c. \biggr] 
     + \frac{1}{2} \lambda_{1} (\phi_1^\dagger \phi_1)^2+\frac{1}{2} \lambda_{2} (\phi_2^\dagger \phi_2)^2 \notag\\
& + \lambda_{3} (\phi_1^\dagger \phi_1)(\phi_2^\dagger \phi_2) 
+ \lambda_{4} (\phi_1^\dagger \phi_2)(\phi_2^\dagger \phi_1)+  \frac{1}{2}\biggl[ \lambda_{5}  (\phi_1^\dagger \phi_2)^2 + h.c. \biggr], 
\end{align}
where $\mu^2_3$ and $\lambda_5$ are complex and the rest of the parameters in the potential are real.
In the presence of an exact $Z_2$ symmetry, using the {\sl rephasing invariance} of \cite{Ginzburg}, 
the phases of the $v_i$'s in Eq.~(\ref{fields}) can be removed by a redefinition of $\mu_3^2$ and $\lambda_5$ and so, henceforth, one can not introduce spontaneous CPV. However, in the case a softly broken $Z_2$ symmetry, spontaneous CPV can occur when $\Im(\lambda^*_5[\mu_3^2]^2)=0$ and there exist no basis in which $\lambda_5$, $\mu_3^2$ and the VEVs are real. 

In this paper, we take the VEVs to be real and positive and study explicit CPV which occurs when $\Im(\lambda^*_5[\mu_3^2]^2) \neq 0$ \cite{Haber:2006ue,Haber:2015pua}. We then define the VEV related to the Fermi constant $G_F$ as $v^2\equiv v_1^2+v_2^2=(\sqrt{2}G_F)^{-1} \simeq (246$ GeV$)^2$ and the ratio of the two VEVs to be $\tan\beta =v_2/v_1$. 
Thus, the only source of CPV in this model is explicit CPV through the complex parameters:
\be 
\mu^2_3=\Re\mu^2_3 + i \Im\mu^2_3, \quad \mbox{and} \quad \lambda_5=\Re\lambda_5+i\Im\lambda_5. 
\ee
In what follows we will be using the notation below
\be 
 \Re\lambda_5 \equiv \lambda_5^r , ~~~~ \Im\lambda_5 \equiv \lambda_5^i. 
\ee

\subsection{Minimising the potential}

The tadpole conditions for the potential,
\be 
\frac{\partial V}{\partial h^0_1}\Big|_0=0,~~
\frac{\partial V}{\partial h^0_2}\Big|_0=0,~~
\frac{\partial V}{\partial a^0_1}\Big|_0=0, 
\ee 
where one gets the same results for $a^0_2$ as for $a^0_1$, lead to the following equations
\bea
&&\mu_1^2-\Re\mu^2_3 ~\tan\beta+\frac{v^2}{2}(\lambda_1 ~c^2_\beta+\lambda_{345}~s^2_\beta)=0, \nonumber\\
&&\mu_2^2-\Re\mu^2_3~\cot\beta+\frac{v^2}{2}(\lambda_2~s^2_\beta)+\lambda_{345}~c^2_\beta)=0, \label{t2}\\
&&\Im\mu_3^2-\frac{v^2}{2}\lambda_{5}^i~ s_\beta ~ c_\beta=0,\nonumber
\eea
where 
\bea
&\lambda_{345}\equiv\lambda_3+\lambda_4+\lambda_5^r. 
\eea
We introduced the abbreviations such that $s_\theta = \sin\theta$,  $c_\theta = \cos\theta$
and $t_\theta = \tan\theta$
 and will use them henceforth. 
Using the first two relations in Eq.~(\ref{t2}), we can eliminate $\mu_1^2$ and $\mu_2^2$ from the potential.
The third relation determines $\Im\mu_3^2$ in terms of other parameters,
\be 
\Im\mu_3^2 = \frac{v^2}{2}\lambda_{5}^i s_\beta  c_\beta.
\ee
Then $\lambda_{5}^i$ may be regarded as the only source of CPV.
We introduce the ``soft breaking scale" of the $Z_2$ symmetry,
\be 
M^2=\frac{\Re\mu_3^2}{s_\beta~c_\beta}.
\label{M-defined}
\ee

It is also useful to introduce the so-called Higgs basis to express the mass matrices for the scalar bosons,  where 
we can separate the Nambu-Goldstone (NG) boson states from the physical ones. In the Higgs basis~\cite{HiggsBasis}, the rotated doublets are represented by $\hat \phi_i$ and are defined as
\be 
\left(\begin{array}{c}
\hat \phi_1 \\ 
\hat \phi_2
\end{array}\right)
= \left(\begin{array}{cc}
c_\beta & s_\beta\\
-s_\beta & c_\beta
\end{array}\right)
\left(\begin{array}{c}
\phi_1 \\ 
\phi_2
\end{array}\right), 
\ee
where
\be 
\hat \phi_1 = \left(
\begin{array}{cc}
$\begin{scriptsize}$ G^+$\end{scriptsize}$\\ 
\frac{v+h'_1+iG^0}{\sqrt{2}}
\end{array}\right), \quad  
\hat \phi_2 = \left(
\begin{array}{cc}
$\begin{scriptsize}$ H^+$\end{scriptsize}$\\ 
\frac{h'_2+ih'_3}{\sqrt{2}}
\end{array}\right), 
\ee
with $G^\pm$ and $G^0$ being the NG bosons absorbed into the longitudinal components of the $W$ and $Z$ bosons, respectively.

The mass of the charged Higgs states, $H^\pm$, is calculated to be
\be 
m^2_{H^\pm}= M^2 -\frac{v^2}{2}(\lambda_4 + \lambda_5^r). 
\ee

The mass matrix for the three neutral states
is given by the $3\times 3$ form in the Higgs basis  ($h_1'$, $h_2'$, $h_3'$) as 
\begin{footnotesize}
\be 
\mathcal{M}^2 =\left(
\begin{array}{ccc}
v^2(\lambda_1c^4_\beta+\lambda_2 s^4_\beta+\frac{1}{2}\lambda_{345}s^2_{2\beta}) 
&
\frac{v^2}{2}s_{2\beta}(\lambda_2s^2_\beta-\lambda_1c^2_\beta+c_{2\beta}\lambda_{345})
&
-\frac{v^2}{2}\lambda_5^i s_{2\beta} 
\\[2mm]
\frac{v^2}{2}s_{2\beta}(\lambda_2s^2_\beta-\lambda_1c^2_\beta+c_{2\beta}\lambda_{345})
&
M^2+v^2 s^2_\beta c^2_\beta(\lambda_1+\lambda_2-2\lambda_{345})
&
-\frac{v^2}{2}\lambda_5^ic_{2\beta}
\\[2mm]
-\frac{v^2}{2}\lambda_5^i s_{2\beta}
&
-\frac{v^2}{2}\lambda_5^i c_{2\beta} 
&
M^2-v^2\lambda_5^r  
\end{array}\right). 
\label{M-squared}
\ee
\end{footnotesize}
This matrix is diagonalised by introducing the $3 \times 3$ orthogonal matrix $R$ as 
\be 
\left(
\begin{array}{c}
h_1'\\
h_2'\\
h_3'
\end{array}\right) = R
\left(
\begin{array}{c}
H_1\\
H_2\\
H_3
\end{array}\right), \quad
R^T\mathcal{M}^2 R =\mathcal{M}^2_{\text{diag}}= \text{diag}(m_{H_1}^2,m_{H_2}^2,m_{H_3}^2), 
\ee
where $H_1$, $H_2$ and $H_3$ represent the mass eigenstates whereas 
$m_{H_1}^2$, $m_{H_2}^2$ and $m_{H_3}^2$ ($m_{H_1}^{}\leq m_{H_2}^{}\leq m_{H_3}^{}$ is assumed by definition) 
are corresponding squared masses. 
In the following, we identify $H_1$ as the SM-like Higgs boson, so that we take $m_{H_1}^{}=125$ GeV, and 
the notations $H_1$ and $h$ will be used interchangeably.

The scalar three point couplings are calculated from the Higgs potential. 
The trilinear neutral Higgs boson couplings  can be extracted in the following way: 
\begin{align}
{\cal L } & =  \lambda_{ijk} h_i'h_j'h_k'  +\cdots \label{lamijk} \\
&=\lambda_{ijk} \sum_{\alpha=1}^3\sum_{\beta=1}^3\sum_{\gamma=1}^3 R_{i\alpha}R_{j\beta}R_{k\gamma}H_\alpha H_\beta H_\gamma +\cdots\notag\\
&=\lambda_{abc}H_aH_bH_c+\cdots, 
\end{align}
where $H_a$ are the mass eigenstates of the neutral Higgs boson and 
\begin{align}
\lambda_{abc} = \sum_{i,j,k=1}^3\lambda_{ijk}[
R_{ia}R_{jb}R_{kc} + (\text{independent permutations of $a$, $b$ and $c$})]. 
\end{align}
The analytic expressions for $\lambda_{ijk}$ and the $H^+H^-H_a$ couplings are given in Appendix~A.

\subsubsection{The $\lambda_5^i =0$ limit}
\label{CPC-limit}

Since $\lambda_5^i$ is the only source of CPV in our model, taking the limit of $\lambda_5^i \to 0$ 
reduces the model to the CPC 2HDM. 
In this limit, the mass matrix for the neutral Higgs bosons, $\mathcal{M}^2$ in Eq.~(\ref{M-squared}), becomes the block-diagonal form
with the $2\times 2$ part and the $1\times 1$ part where the former corresponds to the mass matrix for the CP-even Higgs states 
and the latter to the squared mass of the CP-odd Higgs state. 
The two CP-even states and one CP-odd state can respectively be denoted as $(h,~H)\,(=H_1,~H_2)$ and $A\,(=H_3)$ which is the usual notation in the literature on the CPC 2HDMs. 

The mass matrix for the CP-even Higgs bosons is diagonalised by the angle $\beta-\alpha$ as 
\be 
t_{2(\beta-\alpha)} = \frac{2{\cal M}^2_{12}}{{\cal M}^2_{22}-{\cal M}^2_{11}},  \label{tan2} 
\ee
with the mass squared eigenvalues,
\begin{align}
m_h^2 = {\cal M}^2_{11}s^2_{\beta-\alpha} +  {\cal M}^2_{22}c^2_{\beta-\alpha} - {\cal M}^2_{12}s_{2{(\beta-\alpha)}},  \label{mhsq}\\
m_H^2 = {\cal M}^2_{11}c^2_{\beta-\alpha} +  {\cal M}^2_{22}s^2_{\beta-\alpha} + {\cal M}^2_{12}s_{2{(\beta-\alpha)}}.  \label{mbhsq}
\end{align}
The relation between the Higgs basis $(h_1',h_2')$ and the mass eigenstate basis $(h,H)$ is then given by 
\begin{align}
\begin{pmatrix}
h_1' \\
h_2'
\end{pmatrix} = 
\begin{pmatrix}
s_{\beta-\alpha}  & c_{\beta-\alpha} \\
c_{\beta-\alpha} & -s_{\beta-\alpha}
\end{pmatrix} 
\begin{pmatrix}
h \\
H
\end{pmatrix},  \label{22}
\end{align}
with $0 \leq \beta \leq {\pi}/{2}$.  
The squared mass of $A$ is given by
\begin{align}
m^2_{A} &= {\cal M}^2_{33}. 
\end{align}

\subsubsection{The $\lambda_5^i\ll 1$ case}
\label{CPV-limit}
Note that the parameter $\lambda_5^i$ in Eq.~(\ref{M-squared}), appearing in the off-diagonal elements in the third row and third column, 
is tightly constrained by EDM bounds as they will be discussed in Section~\ref{constraints}. Therefore, we study the model in the
$\lambda_5^i\ll 1$ case where ${\cal M}_{\rm block}^2$ is (upper $2\times2$) block diagonal.
\be 
{\cal R}^T\mathcal{M}^2 {\cal R} = 
\mathcal{M}^2_{\rm block} +~ \mathcal{O}\left(({\lambda_5^i})^2 \right), 
\ee
where the rotation matrix above is
\be
{\cal R} = \begin{pmatrix}
1 & 0 & 0 \\
0 & c_{23} & -s_{23} \\
0 & s_{23} & c_{23} \\
\end{pmatrix}
\begin{pmatrix}
c_{13} & 0 & -s_{13} \\
0 & 1 & 0 \\
s_{13} &0 & c_{13} \\
\end{pmatrix} 
=\begin{pmatrix}
c_{13} & 0 & -s_{13} \\
-s_{13}s_{23} & c_{23} & -c_{13}s_{23} \\
c_{23}s_{13} & s_{23} & c_{13}c_{23} 
\end{pmatrix},
\ee
where $c_{ij}$ and $s_{ij}$ are $\cos(\alpha_{ij})$ and $\sin(\alpha_{ij})$, respectively (with $ij = 13$ or $23$). 
In principle, we allow for 
\be 
-\frac{\pi}{2} < \alpha_{23} \leq \frac{\pi}{2}, \quad 
-\frac{\pi}{2} < \alpha_{13} \leq \frac{\pi}{2}, \quad 
\ee
and the mixing angles can be expressed as
\begin{align}
t_{23}  &=\frac{s_{23}}{c_{23}} =  
\frac{v^2\left({\cal M}^2_{11}-{\cal M}^2_{33}-{\cal M}^2_{12}~t_{2\beta} \right)~\lambda_5^i ~c_{2\beta}}
{2{\cal M}^2_{12}-2\left({\cal M}^2_{11}-{\cal M}^2_{33}\right)\left({\cal M}^2_{22}-{\cal M}^2_{33}\right)} 
+~ \mathcal{O}\left(({\lambda_5^i})^2 \right), \\
t_{13} &=\frac{s_{13}}{c_{13}} = 
\frac{-v^2~c_{23}~s_{2\beta}\lambda_5^i -2 ~c^2_{2\beta}{\cal M}^2_{12} ~s_{23}}
{2 c^2_{2\beta} \left( {\cal M}^2_{11}-{\cal M}^2_{33}~c^2_{23} \right)} + ~ \mathcal{O}\left(({\lambda_5^i})^2 \right). 
\end{align}
Therefore, by neglecting the 
$\mathcal{O}\left(({\lambda_5^i})^2\right)$ contribution, the mass squared matrix is diagonalised by
\bea 
\mathcal{M}_{\text{diag}}^2 &=& R^T\mathcal{M}^2R \notag\\
&=&R^T_{\beta-\alpha} {\cal R}^T \mathcal{M}^2 {\cal R}R_{\beta-\alpha} \notag\\
&\simeq& R^T_{\beta-\alpha}\,  \mathcal{M}^2_{\rm block} \,R_{\beta-\alpha}  , 
\eea
where the upper block is diagonalised in a  similar way to Eq.~(\ref{22}), as
\be 
R_{\beta-\alpha} = \begin{pmatrix}
s_{\beta-\alpha} & c_{\beta-\alpha} & 0 \\
c_{\beta-\alpha} & -s_{\beta-\alpha} & 0 \\
0&0&1
\end{pmatrix} . 
\ee
Using the above expression, we obtain the approximate expression for the diagonalisation matrix $R$:
\begin{align}
R\simeq  
\begin{pmatrix}
s_{\beta-\alpha} & c_{\beta-\alpha} & -s_{13}  \\
c_{\beta-\alpha} & -s_{\beta-\alpha}& -s_{23}  \\
s_{13}+s_{23}c_{\beta-\alpha} & s_{13}c_{\beta-\alpha}-s_{13}s_{\beta-\alpha}   & 1 \
\end{pmatrix}. \label{eq:R}
\end{align}
As described in Subsection \ref{CPC-limit}, we can define the SM-like limit by taking $\lambda_5^i = 0$ 
(equivalently $s_{13}=s_{23}= 0$) and $s_{\beta-\alpha}= 1$, where $H_1$ has the same Yukawa and gauge couplings as those of the SM Higgs boson. 

Therefore, the 9 independent parameters in the model,
\be
\mu_1^2,~
\mu_2^2,~
\text{Re}\,\mu_3^2,~
~\lambda_1,~ \lambda_2,~\lambda_3,~\lambda_4,~\lambda_5^r,~\lambda_5^i.
\ee
can be re-expressed in terms of the following parameters which we shall use as inputs:
\be
v,~\tilde{m}_h,~\tilde{m}_H^{},~\tilde{m}_A^{},~m_{H^\pm}^{},~\tan\beta,~s_{\beta-\tilde{\alpha}},~M^2,~\lambda_5^i, 
\ee
where the parameters with tilde are defined as 
\begin{align}
\tilde{m}_h^2 &\equiv {\cal M}^2_{11}s^2_{\beta-\tilde{\alpha}} +  {\cal M}^2_{22}c^2_{\beta-\tilde{\alpha}} - {\cal M}^2_{12}s_{2{(\beta-\tilde{\alpha})}}, \label{tmhsq} \\
\tilde{m}_H^2 &\equiv {\cal M}^2_{11}c^2_{\beta-\tilde{\alpha}} +  {\cal M}^2_{22}s^2_{\beta-\tilde{\alpha}} + {\cal M}^2_{12}s_{2{(\beta-\tilde{\alpha})}}, \label{tmHsq}\\
t_{2(\beta-\tilde{\alpha})} &\equiv \frac{2{\cal M}^2_{12}}{{\cal M}^2_{22}-{\cal M}^2_{11}}, \label{talpha}\\
 \tilde{m}_A^2 &\equiv {\cal M}_{33}^2 .  
\end{align}
We note that in the CPC limit, 
$\tilde{m}_h$, $\tilde{m}_H^{}$ and $\tilde{m}_A^{}$ correspond to the masses of the two CP-even and 
one CP-odd Higgs bosons, respectively, and 
$\beta-\tilde{\alpha}$ is the mixing angle which diagonalises the CP-even Higgs states in the Higgs basis (see Eqs.~(\ref{tan2}), (\ref{mhsq}) and (\ref{mbhsq})). 

The relation between $m_h(=125~\text{GeV})$ and $\tilde{m}_h$ is described using the parameters defined in Eqs.~(\ref{tmhsq})-(\ref{talpha}) as 
\begin{align}
m_h^2 =  \tilde{m}_h^2c^2_\chi +  \tilde{m}_A^2s^2_\chi 
-  \frac{v^2}{2}\lambda_5^{i}[s_{2\beta} s_{\beta-\tilde{\alpha}} + c_{2\beta}c_{\beta-\tilde{\alpha}}]s_{2\chi}, 
\end{align}
with
\begin{align}
\tan2\chi = \frac{v^2\lambda_5^i }{\tilde{m}_A^2 - \tilde{m}_h^2 }s_{2\beta}. 
\end{align}

In the numerical evaluation, the value of $\tilde{m}_h$ is varied so as to reproduce $125$ GeV.

\section{The Yukawa and kinetic Lagrangian}
\label{yukawa-kinetic}

\begin{table}[t]
\begin{center}
\begin{tabular}{|c||c|c|c|c|c|c||c|c|c|}
\hline & $\Phi_1$ & $\Phi_2$ & $u_R^{}$ & $d_R^{}$ & $e_R^{}$ &
 $Q_L$, $L_L$ &$\xi_u$ & $\xi_d$ & $\xi_e$\\  \hline
Type-I  & $+$ & $-$ & $-$ & $-$ & $-$ & $+$ & $\cot\beta$ & $\cot\beta$ & $\cot\beta$\\
Type-II & $+$ & $-$ & $-$ & $+$ & $+$ & $+$ & $\cot\beta$ & $-\tan\beta$ & $-\tan\beta$\\
Type-X  & $+$ & $-$ & $-$ & $-$ & $+$ & $+$ & $\cot\beta$ & $\cot\beta$ & $-\tan\beta$\\
Type-Y  & $+$ & $-$ & $-$ & $+$ & $-$ & $+$ & $\cot\beta$ & $-\tan\beta$ & $\cot\beta$\\
\hline
\end{tabular} 
\end{center}
\caption{$Z_2$ charge assignment in the four types of Yukawa interactions and the $\xi_f$ factor in each of types.} \label{Tab:type}
\end{table}

The most general form of the Yukawa  Lagrangian
under the introduced $Z_2$ symmetry is given by  
\begin{align}
-{\mathcal L}_Y =
&Y_{u}{\overline Q}_Li\sigma_2\phi^*_uu_R^{}
+Y_{d}{\overline Q}_L\phi_dd_R^{}
+Y_{e}{\overline L}_L\phi_e e_R^{}+\text{h.c.},
\end{align}
where $\phi_{u,d,e}$ are $\phi_1$ or $\phi_2$ 
depending on the type of Yukawa interaction.
When we specify the $Z_2$ charge assignment for fermions as given in Tab.~\ref{Tab:type}, 
$\phi_{u,d,e}$ are determined. 
For example, in the Type-II 2HDM $\phi_d=\phi_e=\phi_1$ and $\phi_u=\phi_2$. 
The interaction terms  are expressed as
\begin{align}
-{\mathcal L}_Y^{\text{int}}=&
\sum_{f=u,d,e}\frac{m_f}{v}\sum_{i=1,2,3}\left(\xi_f^{H_i}{\overline f}fH_i-2i\,I_f \tilde{\xi}_f^{H_i}   {\overline f} \gamma_5 fH_i\right)\notag\\
&+\frac{\sqrt{2}}{v}\Big[V_{ud}\overline{u}
\left(m_d\xi_d\,P_R-m_u\xi_uP_L\right)d\,H^+
+m_e\xi_e\overline{\nu^{}}P_Re^{}H^+
+\text{h.c.}\Big],  \label{yukawa_thdm}
\end{align}
where $I_f$ is the third component of the isospin for a fermion $f$ and the $\xi_f$ values 
are listed in Tab.~\ref{Tab:type}. 
In Eq.~(\ref{yukawa_thdm}),  
the coefficients for the scalar (pseudo-scalar) type couplings 
$\xi_f^{H_i}$ $(\tilde{\xi}_f^{H_i})$ are given by
\begin{align}
\xi_f^{H_1}&=R_{11}+\xi_f R_{21} 
\simeq s_{\beta-\alpha}+\xi_f c_{\beta-\alpha}, \\
\xi_f^{H_2}&=R_{12}+\xi_f R_{22} 
\simeq c_{\beta-\alpha}-\xi_f s_{\beta-\alpha},  \\
\xi_f^{H_3}&=R_{13}+\xi_f R_{23}  
\simeq  -s_{13} - s_{23}\xi_f , \\
\tilde{\xi}_f^{H_1}&=\xi_f R_{31} \simeq  \xi_f (s_{13}+s_{23}c_{\beta-\alpha}) , \label{xit1}  \\
\tilde{\xi}_f^{H_2}&= \xi_f R_{32} \simeq  \xi_f (s_{13}c_{\beta-\alpha}-s_{13}s_{\beta-\alpha} ), \\
\tilde{\xi}_f^{H_3}&= \xi_f R_{33} \simeq  \xi_f, 
\end{align}
where the approximated formulae given in the above rightmost hand sides are obtained using Eq.~(\ref{eq:R}) which is valid for the case of $\lambda_{5}^i\ll 1$. 

The kinetic terms for the scalar fields are given by 
\begin{align}
{\mathcal L}_{\text{kin}}& =|D_\mu \phi_1|^2+|D_\mu \phi_2|^2=|D_\mu \hat \phi_1|^2+|D_\mu \hat \phi|^2. \label{kin}
\end{align}
The gauge-gauge-scalar type interactions only appear from the first, $|D_\mu \hat \phi_1|^2$. They are extracted as
\begin{align}
|D_\mu \hat \phi_1|^2 
&= g_{hVV}^{\text{SM}}(\xi_V^{H_1}H_1+\xi_V^{H_2} H_2 + \xi_V^{H_3}H_3)V_\mu V^\mu  + \cdots ,\quad V_\mu=W_\mu~,Z_\mu, 
\end{align}
where $g_{hVV}^{\text{SM}}$ is the $hVV$ vertex in the SM, and 
\begin{align}
&\xi_V^{H_1} = R_{11}\simeq s_{\beta-\alpha}, \label{xi1}\\
&\xi_V^{H_2} = R_{12}\simeq c_{\beta-\alpha}, \label{xi2}\\
&\xi_V^{H_3} = R_{13}\simeq -s_{\beta-\alpha}s_{13}+c_{\beta-\alpha} s_{23}. \label{xi3}
\end{align}

Note that the alignment limit in which the coupling of $H_1\,(=h)$ are exactly SM-like is achieved in the limit of  
 $\lambda_{5}^{i}\to 0$ (equivalently $s_{13}=s_{23}=0$) and $s_{\beta-\alpha} \to 1$.

Similar to the discussion of the Yukawa couplings, the approximated formulae given in the above rightmost hand sides are obtained using Eq.~(\ref{eq:R}). 
The scalar-scalar-gauge type interactions are also extracted from Eq.~(\ref{kin}):
\begin{align}
|D_\mu \hat \phi_2|^2 & =  
 -\frac{g}{2}\Big[(R_{31} + iR_{21})H^+ \overleftrightarrow{\partial}_\mu H_1  
+(R_{32} + iR_{22})H^+ \overleftrightarrow{\partial}_\mu H_2 \notag\\
&\hspace{13mm}+(R_{33} + iR_{23})H^+ \overleftrightarrow{\partial}_\mu H_3 \Big]W^{-\mu}+ \text{h.c.}\notag\\
&+ \frac{g_Z^{}}{2}\Big[(R_{21}R_{32}+R_{22}R_{31})H_1 \overleftrightarrow{\partial}_\mu H_2
+(R_{21}R_{33}+R_{23}R_{31})H_1 \overleftrightarrow{\partial}_\mu H_3\notag\\
&\hspace{13mm} +(R_{22}R_{33}+R_{23}R_{32})H_2 \overleftrightarrow{\partial}_\mu H_3\Big]Z^\mu +\cdots , 
\end{align}
where $X\overleftrightarrow{\partial}_\mu Y \equiv X(\partial_\mu Y) -Y(\partial_\mu X) $.

\section{Numerical results in the Type-I 2HDM with CPV}
\label{plots}

\subsection{Constraints on the parameters}
\label{constraints}

\subsubsection{Theoretical bounds}
The stability condition for the Higgs potential is given by requiring that 
the potential be bounded from below in any direction of the scalar boson space. 
The necessary and sufficient conditions to guarantee such a positivity of the potential are \cite{3hdm_vs}
\be
\lambda_1 > 0,\quad \lambda_2 > 0, \quad \sqrt{\lambda_1\lambda_2} + \lambda_3 + \text{MIN}(0,~\lambda_4-|\lambda_5|) >0.   \label{vs3}  
\ee

From the S-matrix unitarity for elastic scattering of 2 body to 2 body bosonic states, the magnitude of combinations of $\lambda$ parameters
in the potential can be constrained. 
In Refs.~\cite{PU-CPC1,PU-CPC2}, 
the diagonalised $s$-wave amplitude matrix for these scattering processes has been derived in the CPC 2HDM. 
For the CPV case, we obtain all the eigenvalues of the $s$-wave amplitude matrix just 
by replacing $\lambda_5^r$ with  $|\lambda_5| = \sqrt{(\lambda_5^r)^2 + (\lambda_5^i)^2}$~\cite{Ginzburg:2005dt,PU-CPV2}. 

As for the constraints from experimental data, 
we take into account EDMs and the $S$, $T$ and $U$ parameters~\cite{Peskin-Takeuchi,STU-THDM}. 
In particular, the CPV parameter, i.e., $\lambda_5^i$ 
can significantly affect  EDMs, so its magnitude is constrained. 
The bounds from the EDM constraints have been discussed in Refs.~\cite{EDM1,EDM2} in  CPV 2HDMs. 
In general, there are two sources which contribute to EDMs in  CPV 2HDMs, namely, 
the modified couplings of the SM-like Higgs boson and contributions from additional Higgs bosons. 
In the Type-I 2HDM, 
the pseudo-scalar type interaction among the additional Higgs bosons and fermions are suppressed by 
the factor of $1/\tan\beta$ as we see Eq.~(\ref{xit1}) with $\xi_u=\xi_d=\xi_e = \cot\beta$, so that 
the additional Higgs boson contributions can be neglected in a large $\tan\beta$ regime. 
In the following, we focus on the Type-I 2HDM and we apply the bound from EDMs in the following way~\cite{EDM2}
\begin{align}
\tilde{\xi}_u^{H_1} \leq 10^{-2}.  \label{edm}
\end{align}
Regarding  the $S$, $T$ and $U$ parameters, 
we use the following bounds~\cite{Baak:2012kk} on the deviations in these parameters under the fixed value of $\Delta U = 0$:
\begin{align}
\Delta S =0.05\pm 0.09,\quad  \Delta T = 0.08\pm 0.07,
\label{STallowed}
\end{align}
where $\Delta X$ is the difference between the $X=(S,~T~\text{or}~U)$ parameter in the 2HDM and  in the SM. 
The correlation coefficient of $\Delta S$ and $\Delta T$ is taken  to be $+0.91$.

\subsubsection{Experimental bounds}

The $B$ physics data also provides constraints on the parameter space in 2HDMs, which are especially sensitive to 
$m_{H^\pm}$ and $\tan\beta$. 
A comprehensive study for the constraint on the CPC 2HDMs has been done in Ref.~\cite{stal}, where various 
$B$ physics observables such as $b\to s\gamma$, $B^0$-$\bar{B}^0$ mixing, $B \to \tau\nu$ have been taken into account. 
In the CPV 2HDM, the Yukawa couplings of the charged Higgs boson are the same as those of the CPC 2HDMs, therefore
we can apply the same bound related to the $H^\pm$ mediation as that reported in \cite{stal} to the CPV case studied here\footnote{In Ref.~\cite{BaBar}, the BaBar Collaboration has reported that the measured ratios $\text{BR}(B \to D^{*} \tau \nu) / \text{BR}(B \to D^{*} \ell \nu)$ and $\text{BR}(B \to D \tau \nu) / \text{BR}(B \to D \ell \nu)$ ($\ell = e, \mu$) 
deviate from the SM predictions by $2.7\sigma$ and $2.0\sigma$, respectively, and their combined deviation is $3.4\sigma$.  
These deviations cannot be simultaneously compensated by a natural flavor conserving version such as a $Z_2$ symmetric 2HDMs with and without CPV. 
}.

In addition, we also take into account the constraint from direct searches for extra Higgs bosons at the LHC. 
The search for neutral Higgs bosons decaying into $\tau\tau$ 
using the LHC Run-I data reported in \cite{tautau}, excludes $\tan\beta \gtrsim 10~(30)$ for $m_A=300$ (700) GeV in the minimal supersymmetric SM.
A similar bound is expected in the non-supersymmetric Type-II 2HDM, since the structure of the Yukawa interactions are the same. 
However, there is no $\tan\beta$ enhancement in the Yukawa couplings in the Type-I 2HDM studied here since the Yukawa couplings are suppressed by the factor of $\cot\beta$. The production cross section is, therefore, suppressed by $\cot^2\beta$. 
As a result, since we do not consider the case of $\tan\beta \ll 1$, our model satisfies the constraint from the direct searches at the LHC.

There are also constraints from the $A \to Zh$ process \cite{AZh} which we need to take into account.
The upper limit on the $\sigma(gg \to A)\times \text{BR}(A \to Zh) \times \text{BR}(h \to f\bar{f})$
has been given in the region of $m_A=220$-1000 GeV using the LHC Run-I data. 
For $f=\tau\,(b)$, the upper limit is measured to be $0.098-0.013$ pb ($0.57-0.014$ pb). 
In our model, the typical cross section of $gg \to H_{2,3}$ is of order $1$ pb in the case of $m_{H_{2,3}}=200$ GeV and $\tan\beta \gtrsim 2$, and 
the branching fraction of the $A\to Zh$ mode is less than order of $10^{-2}$. 
On the other hand, the decay rate of the SM-like Higgs boson does not change so much from the SM prediction, so that 
the branching fraction of $h \to \tau\tau (b\bar{b})$ is $\sim 7\% (60\%)$. 
Therefore, our prediction of the cross section is well below the upper limit.

\begin{figure}[h!]
\begin{center}
\includegraphics[scale=0.32]{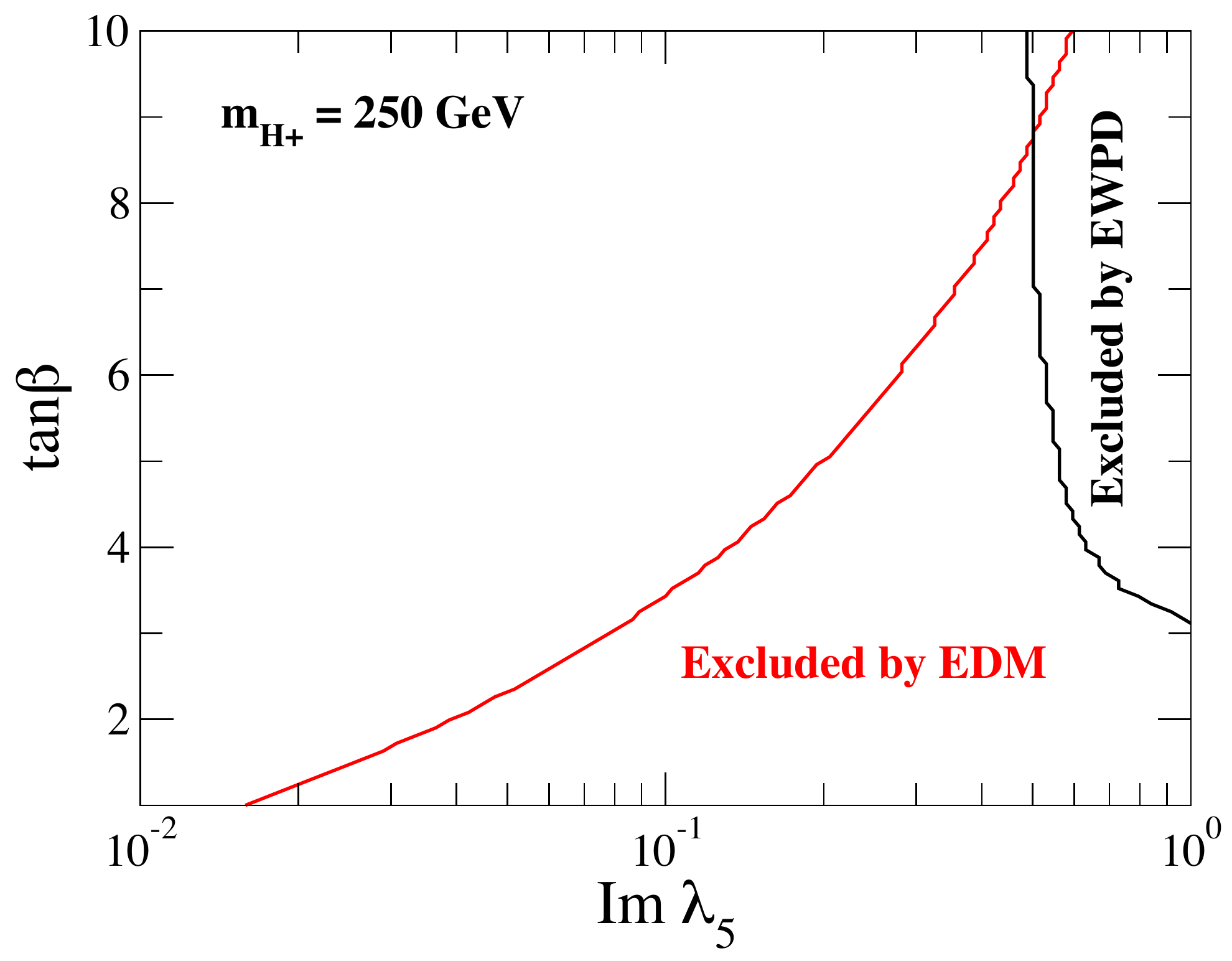} \hspace{3mm} 
\includegraphics[scale=0.32]{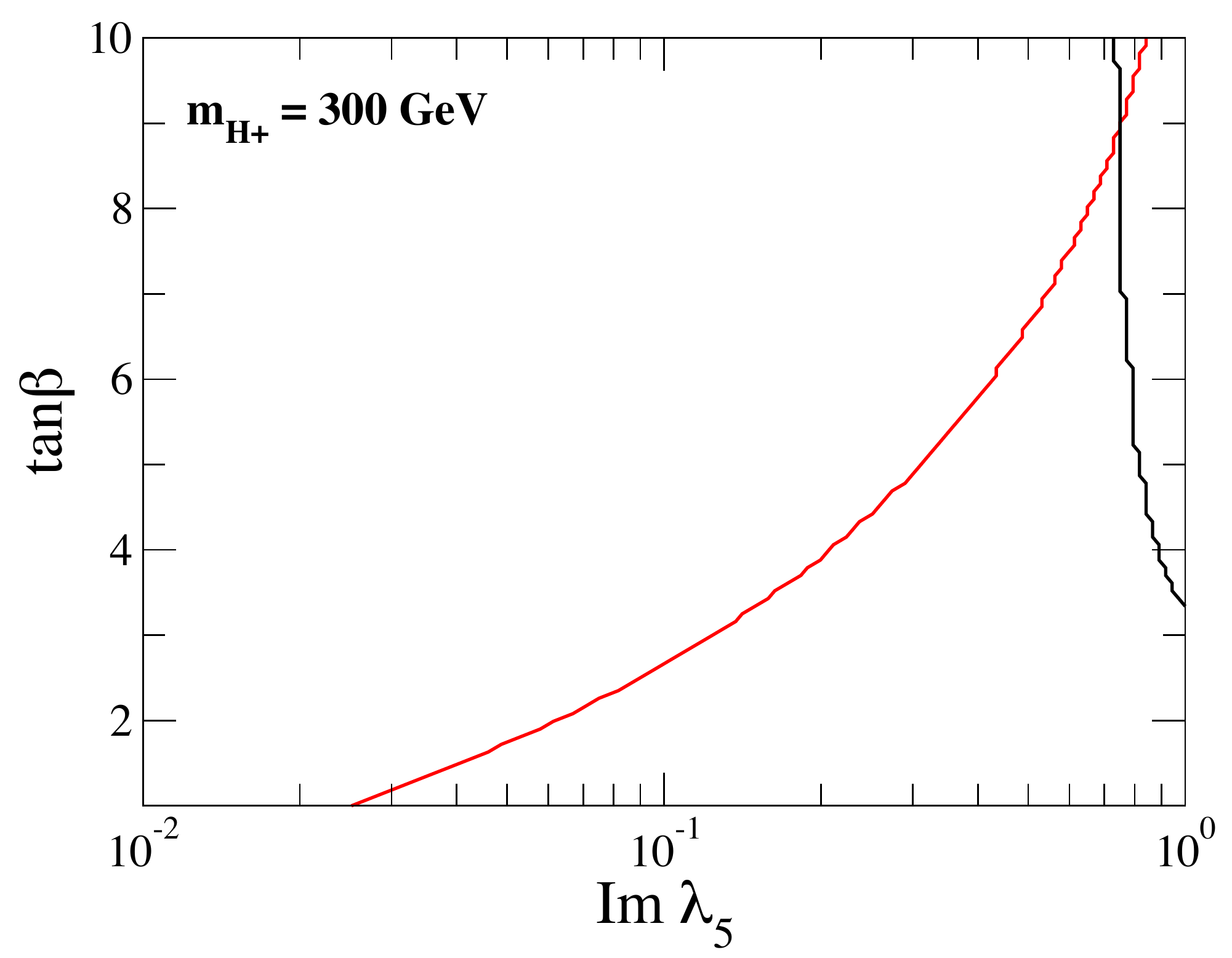} \hspace{3mm} 
\includegraphics[scale=0.32]{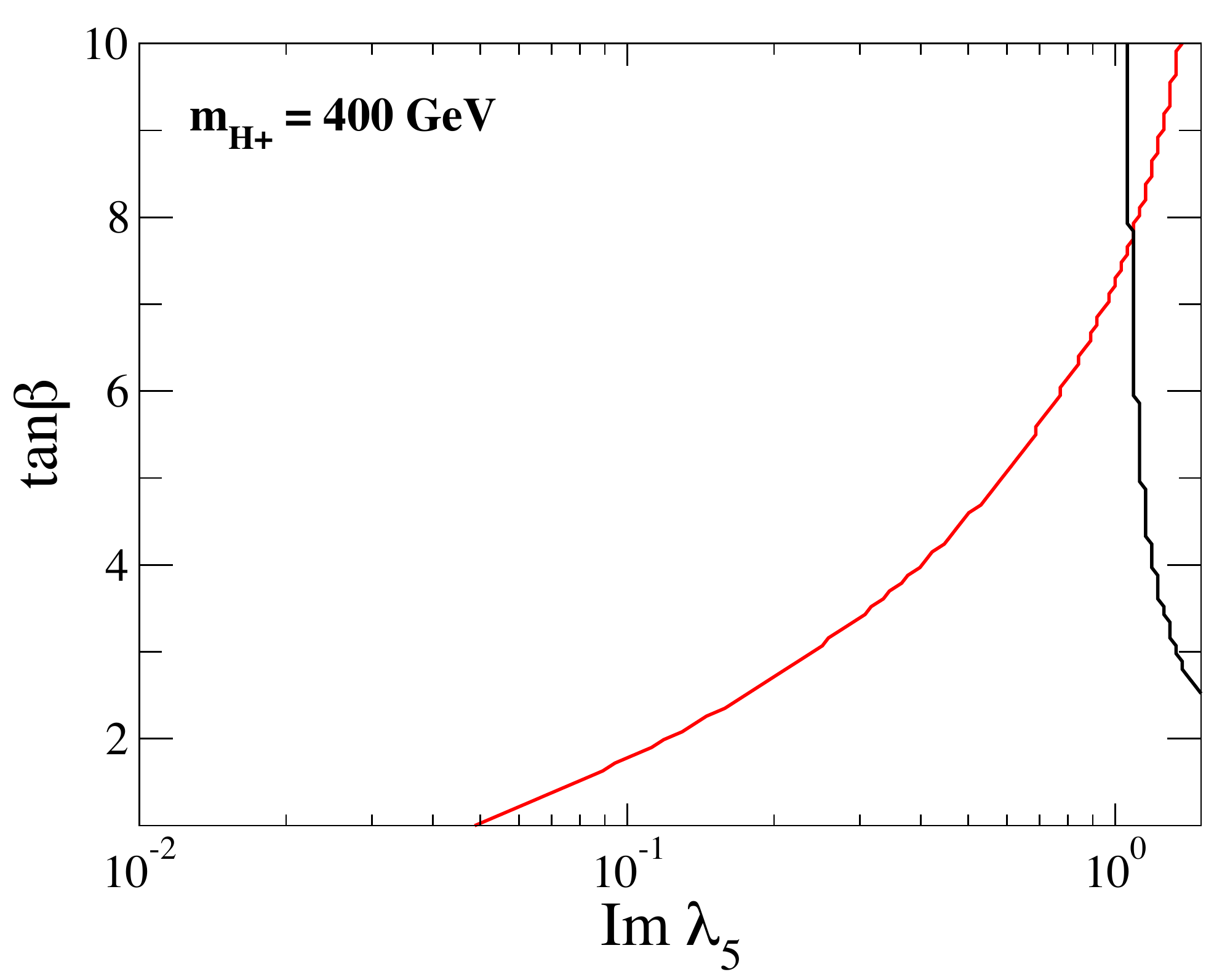} \hspace{3mm} 
\includegraphics[scale=0.32]{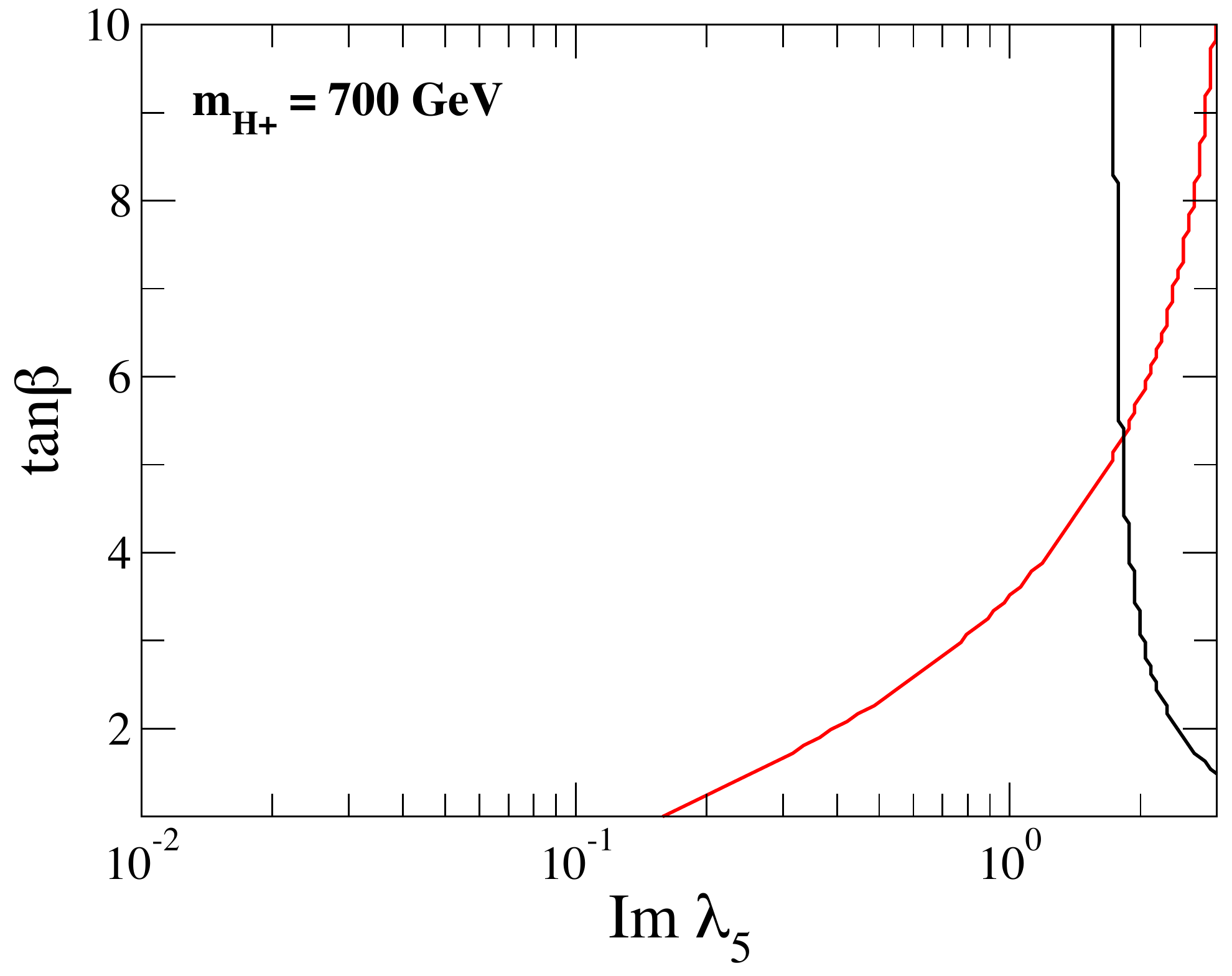} \hspace{3mm} 
\caption{The constrained region in the $\lambda_5^i$-$\tan\beta$ plane is shown in the case of 
$\tilde{m}_H^{}$ = 200 GeV, $\tilde{m}_A^{} = m_{H^\pm}^{}$ and $s_{\beta-\tilde{\alpha}}=1$. 
The upper-left, upper-right, lower-left and lower-right panels respectively show the case of 
$m_{H^\pm}=250$, 300, 400 and 700 GeV. 
For all the panels, 
the right regions from the red and black curves are excluded by the EDM and the electroweak $S$ and $T$ parameters bounds, respectively. }
\label{bounds1}
\end{center}
\end{figure}

\begin{figure}[h!]
\begin{center}
\includegraphics[scale=0.32]{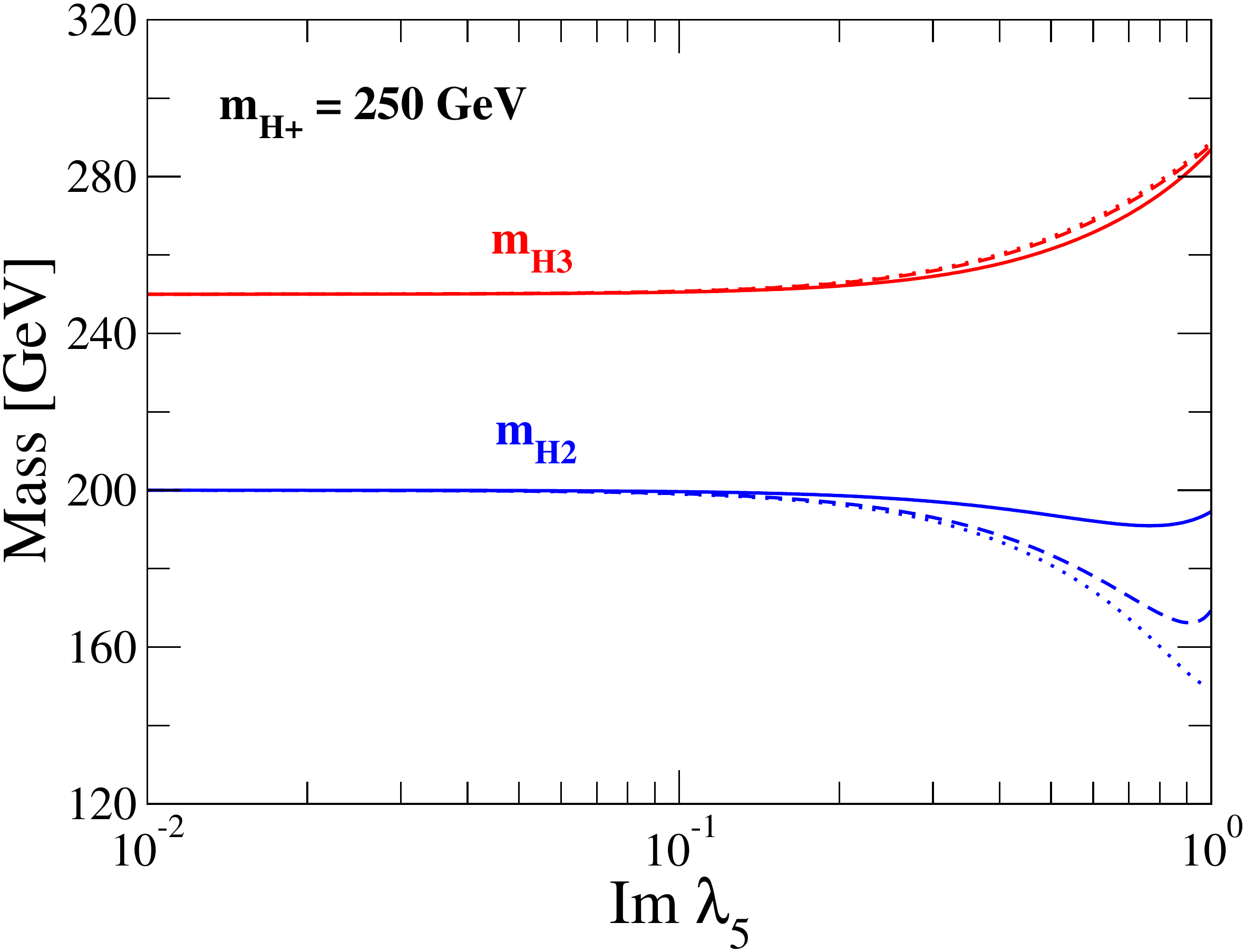}
\includegraphics[scale=0.32]{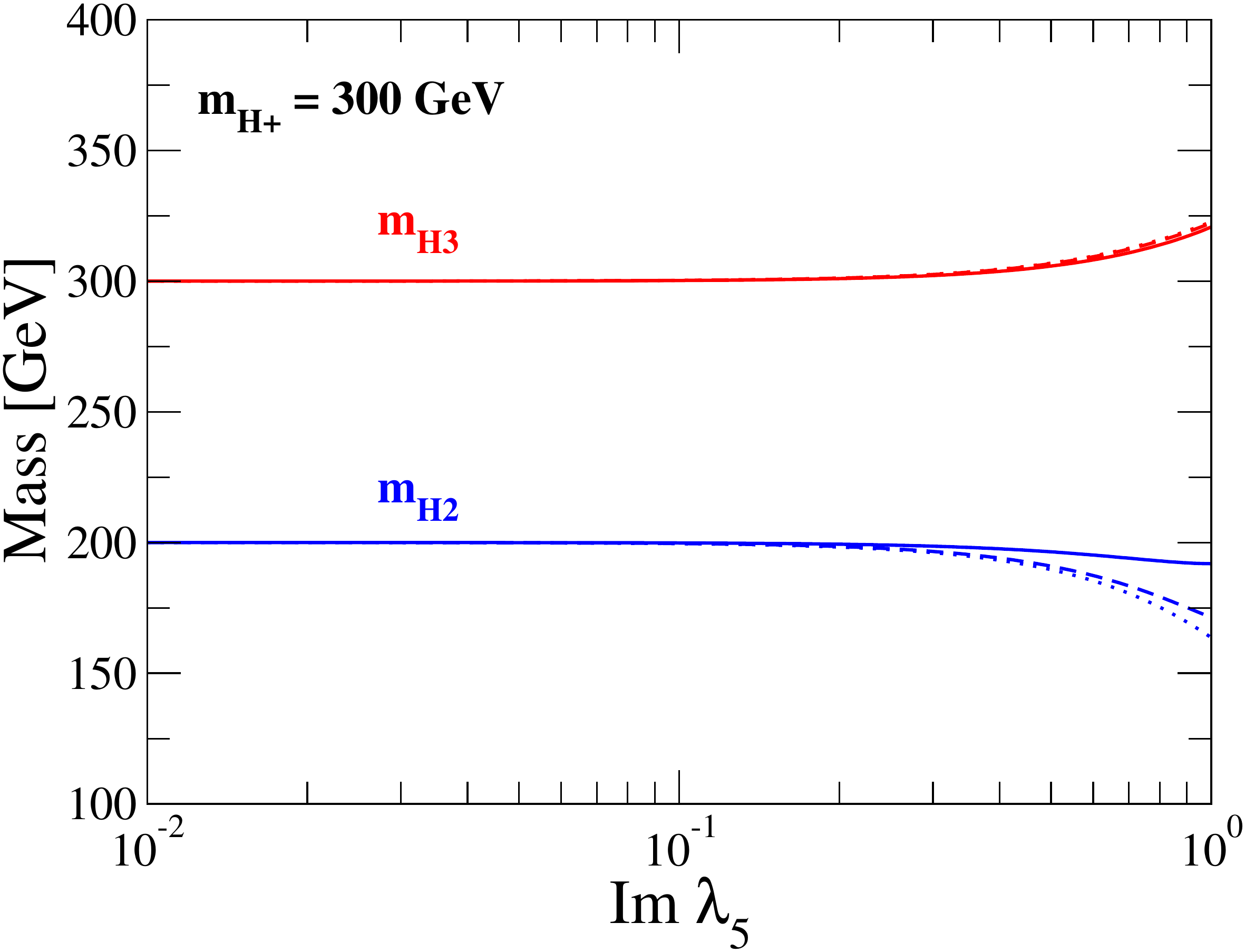}\\
\includegraphics[scale=0.32]{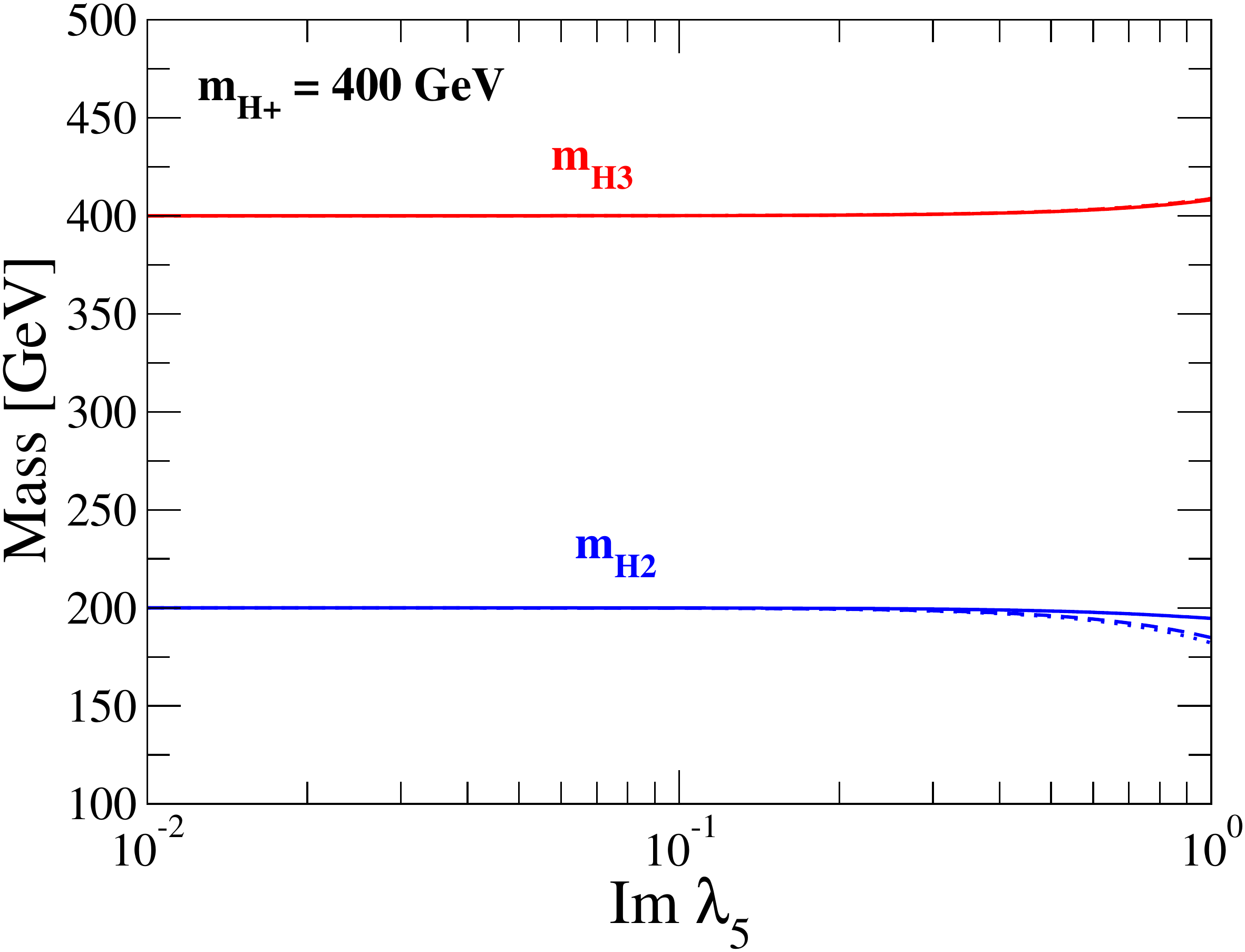}
\includegraphics[scale=0.32]{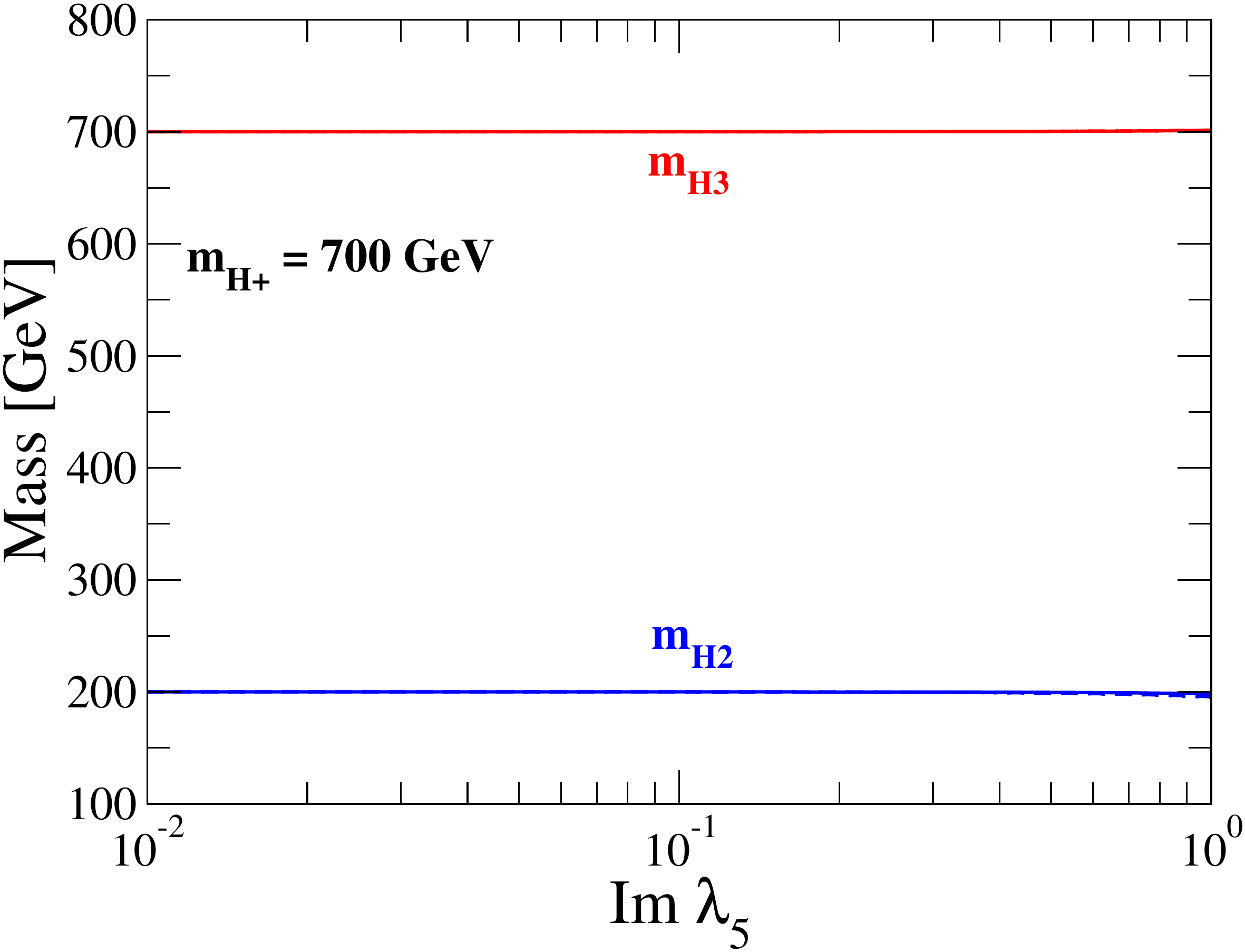}
\caption{The masses of $H_2$ and $H_3$ as a function of $\lambda_5^i$. 
We take the same parameter set as in Fig.~\ref{bounds1}. The mass of the SM-like Higgs boson $H_1$ is kept to be 125 GeV. In each plot the solid, dashed and dotted curves correspond to $\tan\beta=2$, $5$ and $10$, respectively.
}
\label{masses}
\end{center}
\end{figure}

\begin{figure}[h!]
\begin{center}
\includegraphics[scale=0.182]{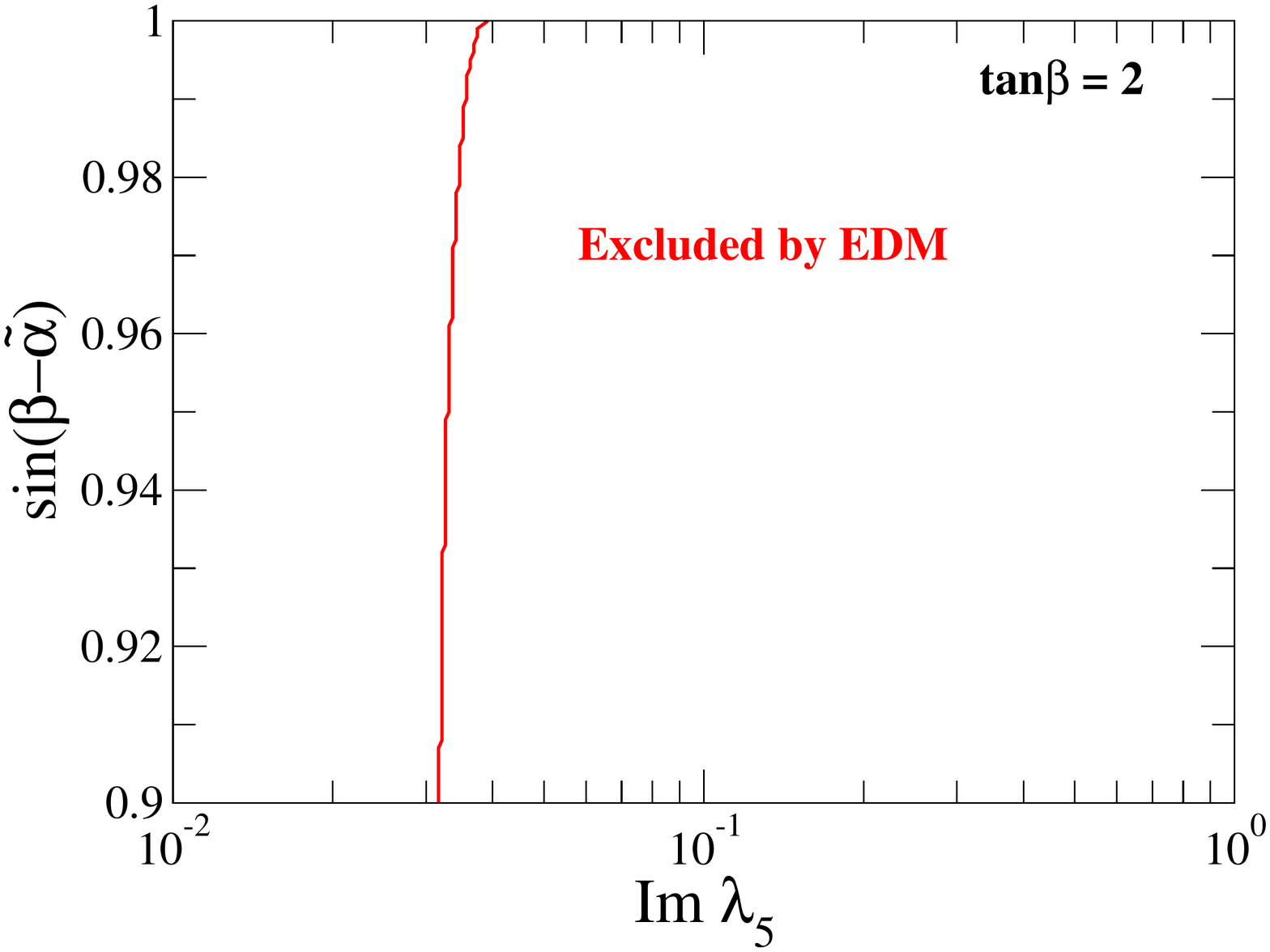}
\includegraphics[scale=0.182]{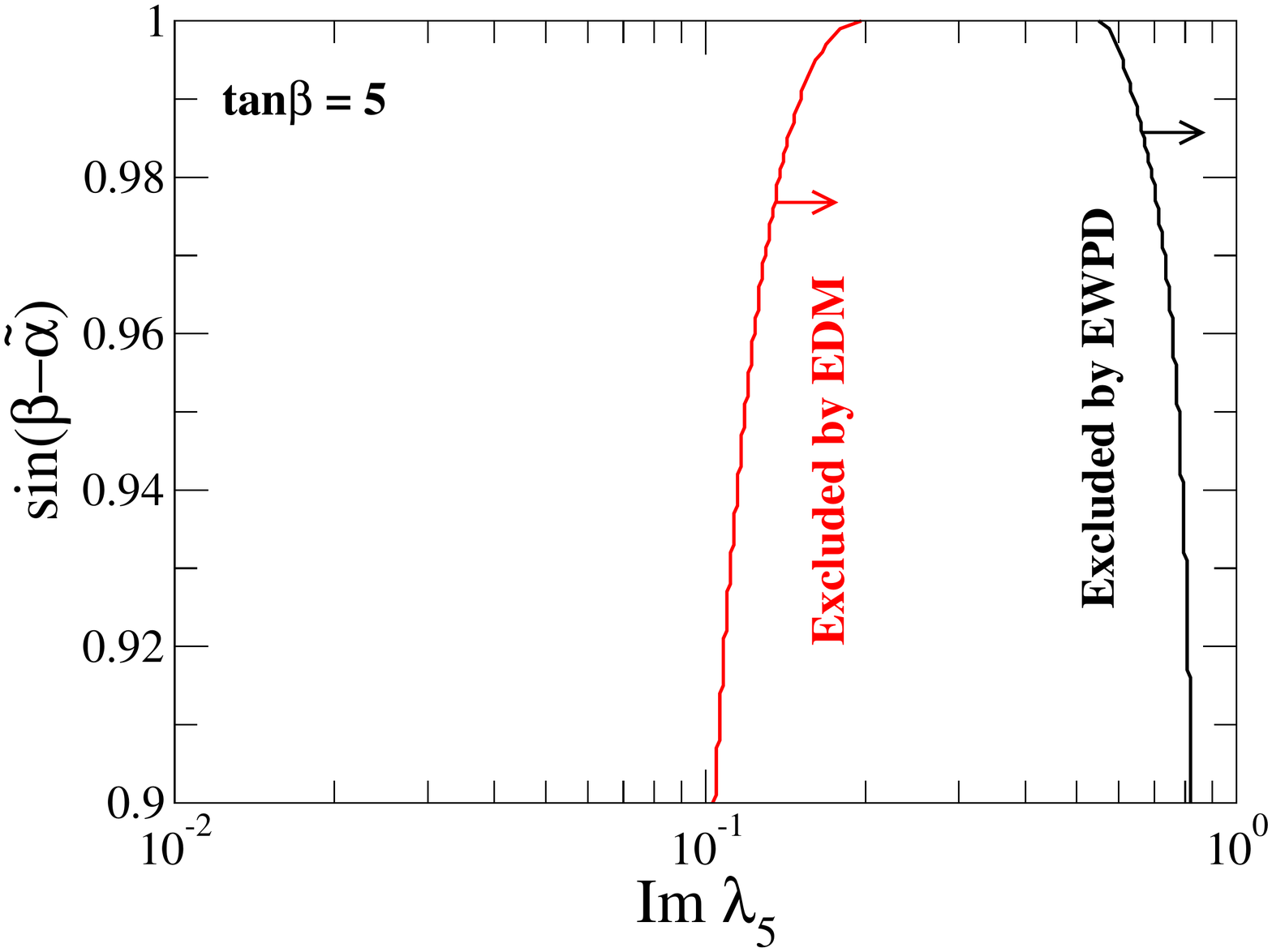}
\includegraphics[scale=0.182]{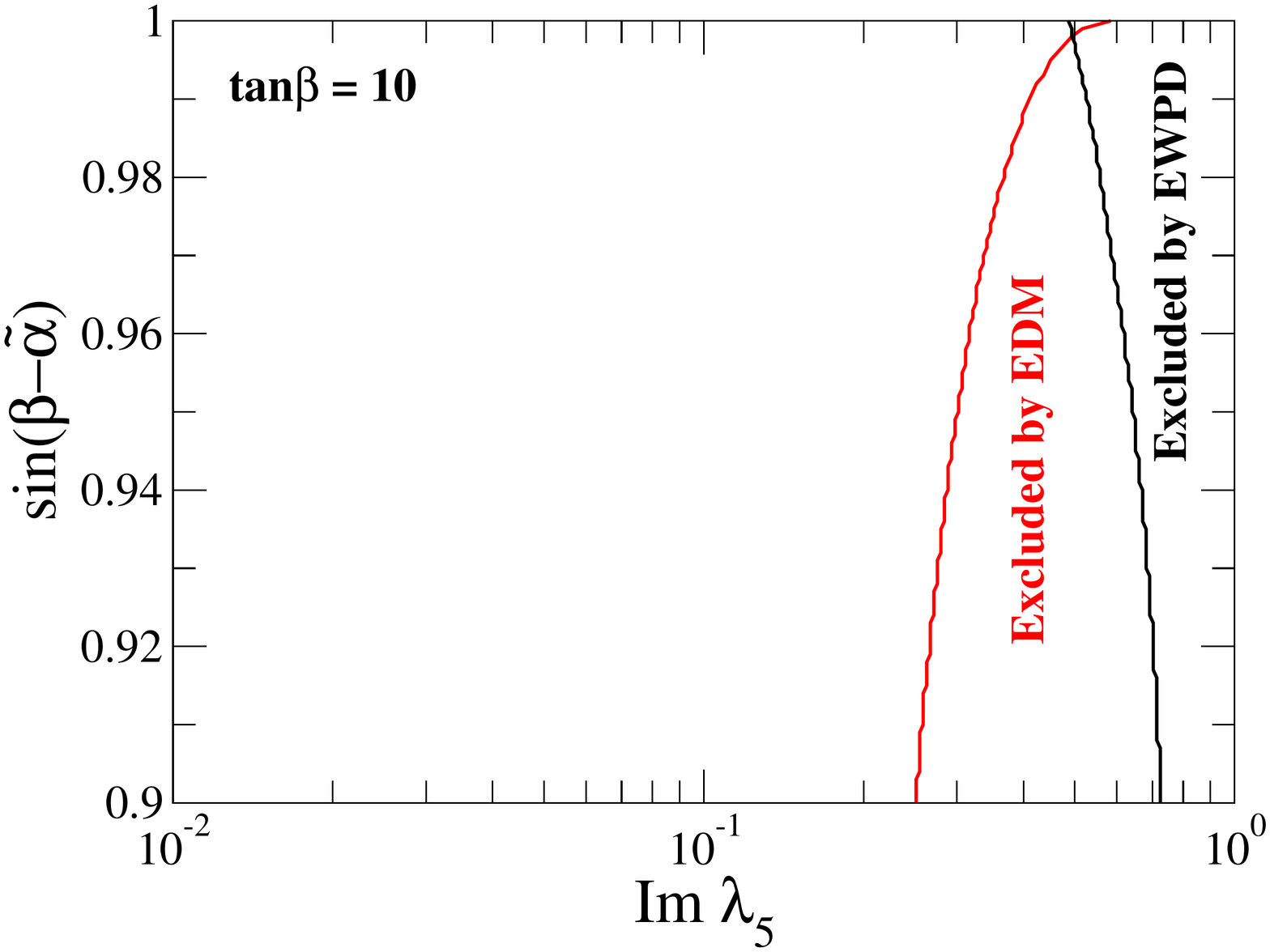}
\caption{The constrained region in the $\lambda_5^i$-$s_{\beta-\tilde{\alpha}}$ plane is shown in the case of 
$\tilde{m}_H^{}$ = 200 GeV and $\tilde{m}_A^{} = m_{H^\pm}^{}=250$ GeV. 
The left, center and right panels  show the case of $\tan\beta=2$, 5 and 10, respectively. 
For all the panels, the right regions from the red and black curves are excluded by the EDMs and the $S$ and $T$ parameters bounds, respectively. }
\label{bounds3}
\end{center}
\end{figure}

In Fig.~\ref{bounds1}, we show the allowed parameter regions 
on the $\lambda_5^i$ and $\tan\beta$ plane from the EDMs given by Eq.~(\ref{edm}) and the $S$ and $T$ parameters given by Eq.~(\ref{STallowed}). 
We take $\tilde{m}_H^{}$ = 200 GeV, $\tilde{m}_A^{} = m_{H^\pm}^{}$ and $s_{\beta-\tilde{\alpha}}=1$. 
The mass of the charged Higgs boson $m_{H^\pm}$ is taken to be 250, 300, 400 and 700 GeV. 
We note that the bounds from the EDMs and the $S$ and $T$ parameters
do not depend on the value of $M^2$. 
Although the $M^2$ dependence appears in the constraints from the unitarity and vacuum stability, 
these bounds can be avoided by taking an appropriate value of $M^2$ for each fixed value of $\tan\beta$ and $\lambda_5^i$. 
We confirmed that the case for $m_{H^\pm}\gtrsim$ 750 GeV is excluded by unitarity bounds\footnote{ 
Note that this upper limit on $m_{H^\pm}$ is due to the assumption that the masses of other scalars are relatively close. If one takes the decoupling limit into account, the mass of the charged scalar could be arbitrarily high without violating any unitarity limits.}.

Because the masses of neutral Higgs bosons are derived as output, we show $m_{H_2}^{}$ and $m_{H_3}^{}$ as a function of $\lambda_5^i$ in Fig.~\ref{masses}. 
As we explained in Subsection~\ref{CPV-limit},  the mass of the SM-like Higgs boson $m_{H_1}^{}$ is kept to be 125 GeV by taking an appropriate value of $\tilde{m}_{h}^{}$ 
for each fixed values of the input parameters. In this figure, we take the same set of input parameters as in Fig.~\ref{bounds1}. 
We see that for the case with $\lambda_5^i\lesssim 0.1$, 
$m_{H_2}^{}\simeq \tilde{m}_{H}$ and $m_{H_3}^{}\simeq \tilde{m}_A$ are given. 
However, when we take a larger value of $\lambda_5^i$, the above approximate relations are broken due to the CP-mixing effect. 
This behaviour is getting more significant when we take a smaller value of $m_{H^\pm}^{}$. 
As it will become clear later, what  is important to note now is the fact that $m_H{_2}$ and $m_{H_3}$ are never degenerate.

In Fig.~\ref{bounds3}, we show the excluded parameter space due to EDMs and the $S$ and $T$ parameters in the $\lambda_5^i$-$s_{\beta-\tilde{\alpha}}$ plane 
for different values of $\tan\beta$, namely, $\tan\beta = 2$ (left panel), 5 (center panel) and 10 (right panel).
In these plots, we take $\tilde{m}_{H}=200$ GeV and $\tilde{m}_{A}^{} = m_{H^\pm}^{}=250$ GeV. 

\subsection{Phenomenology at the LHC}

\begin{figure}[h!]
\begin{center}
\includegraphics[scale=0.26]{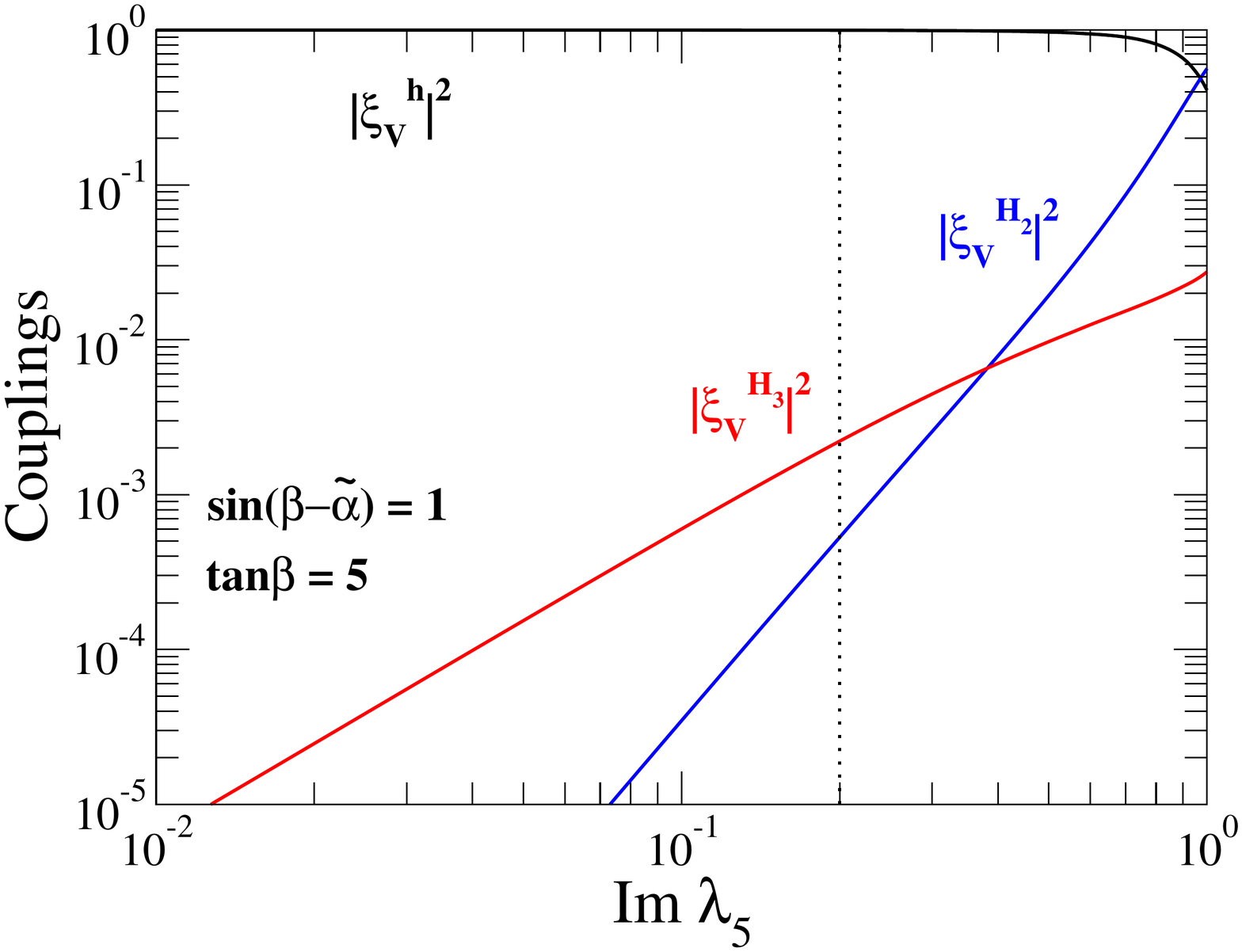} \hspace{5mm}
\includegraphics[scale=0.26]{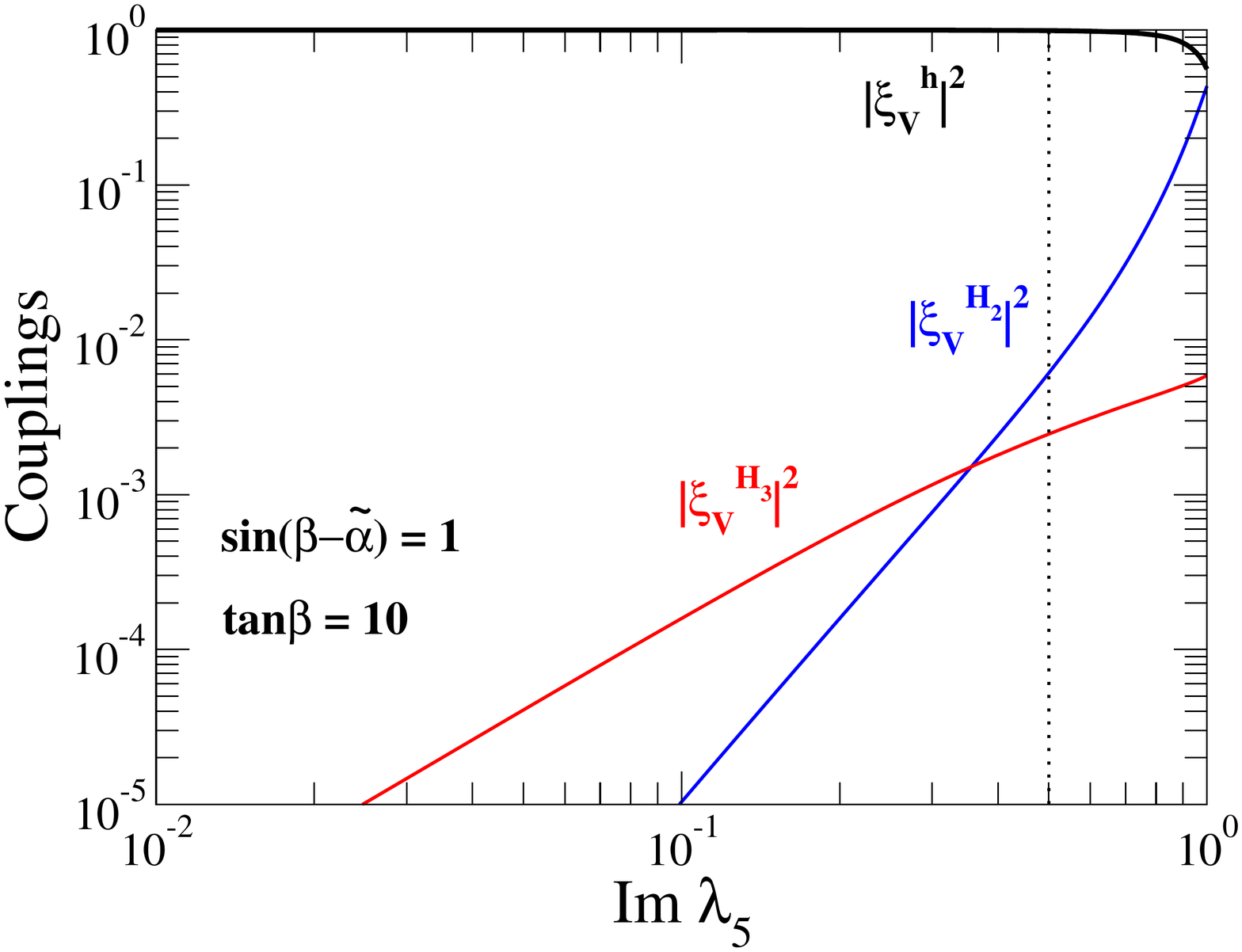} \\
\includegraphics[scale=0.26]{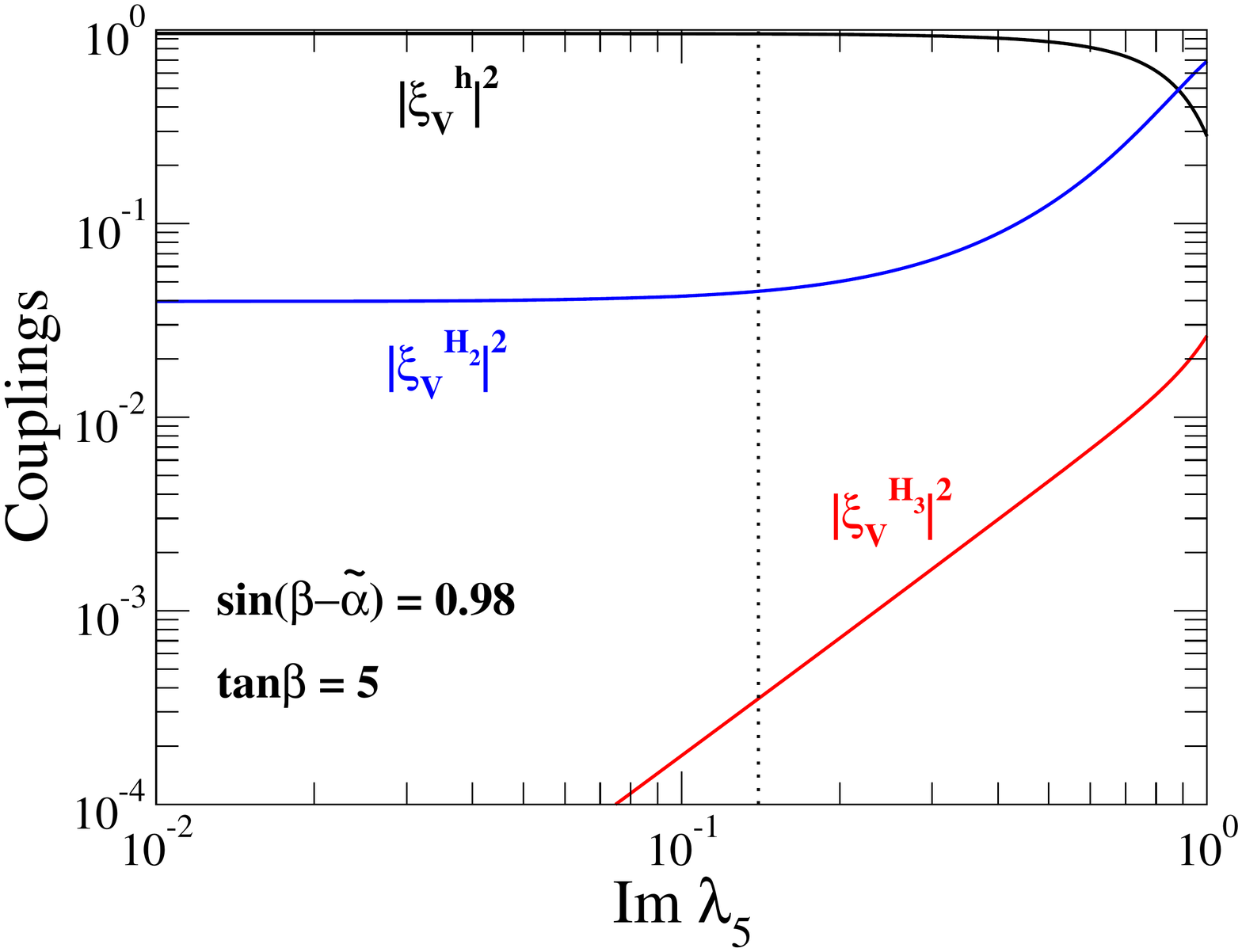}  \hspace{5mm}
\includegraphics[scale=0.26]{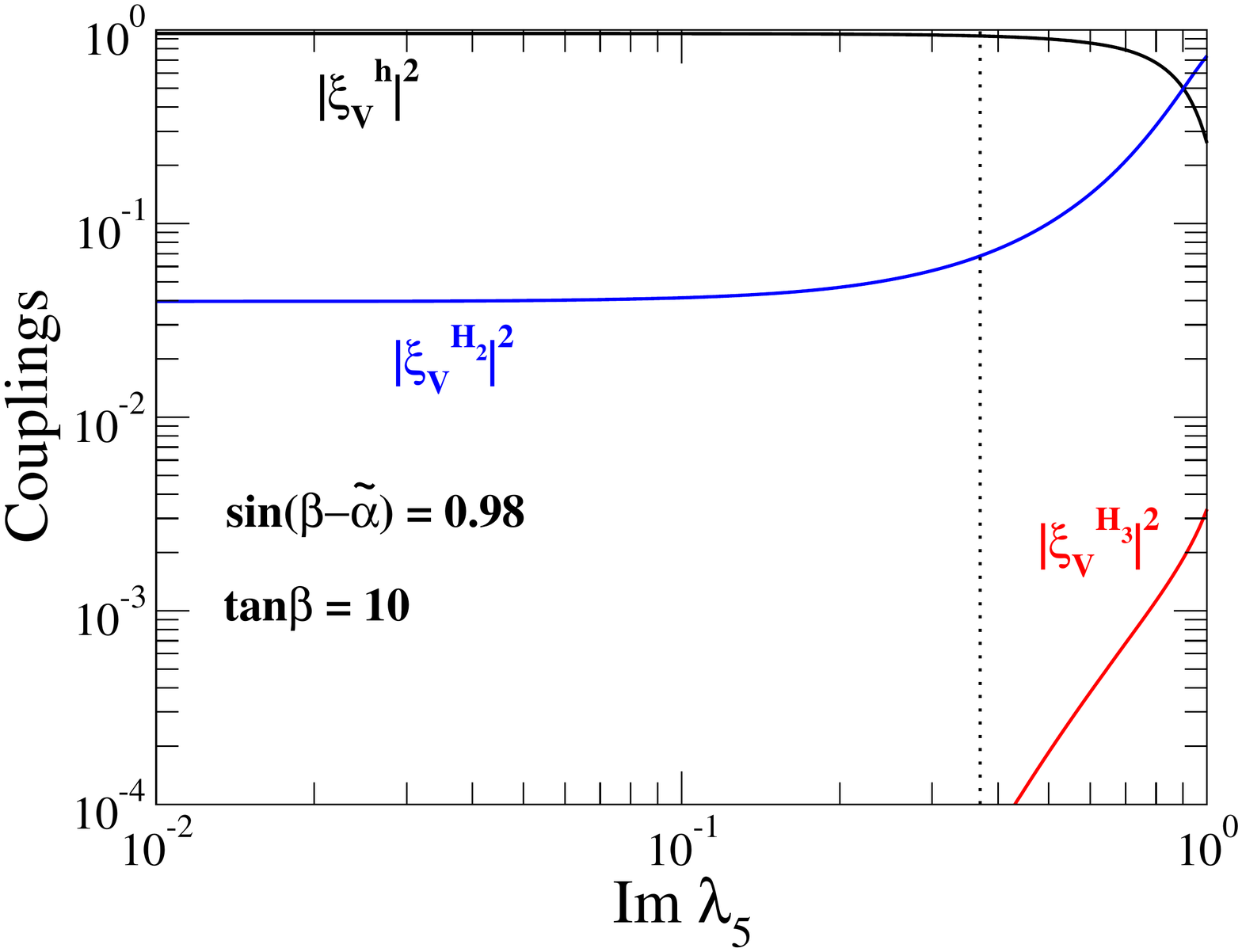} 
\caption{The coefficient of the gauge-gauge-scalar type couplings 
for $h(=H_1)$, $H_2$ and $H_3$ 
defined in Eqs.~(\ref{xi1}), (\ref{xi2}) and (\ref{xi3}), respectively,
as a function of $\lambda_5^i$ for $\tan\beta = 5$ (left) and  $\tan\beta = 10$ (right). 
The value of $s_{\beta-\tilde{\alpha}}$ is taken to be 1 in the upper panels and 0.98 in the lower panels. 
For all the plots, we take $\tilde{m}_{H}=200$ GeV and $\tilde{m}_A^{}=m_{H^\pm}^{}=250$ GeV.  
The vertical dotted line shows the upper limit on $\lambda_5^i$ from the EDMs and $S$ and $T$ parameters. }
\label{gauge}
\end{center}
\end{figure}

For our numerical results, we use the fixed input parameters $\tilde{m}_H^{}=200$ GeV and $\tilde{m}_{A}^{}=m_{H^\pm}^{}=250$ GeV which correspond to the 
case shown in the upper-left panel of Figs.~\ref{bounds1}--\ref{masses} and in Fig.~\ref{bounds3}. 

For the calculations of decay rates of the Higgs bosons, 
it is important to show the value of gauge-gauge-scalar type couplings which are described by 
$g_{hVV}^{\text{SM}}\times \xi_V^{H_i}$ ($i=1,2,3$) given in Eqs.~(\ref{xi1})--(\ref{xi3}). 
We thus first show the values of $\xi^{H_i}_V$ as a function of $\lambda_5^i$ in Fig.~\ref{gauge}. 
In this plot, $\tan\beta$ is fixed to be 5 (left panels) and 10 (right panels). 
The value of $s_{\beta-\tilde{\alpha}}$ is taken to be $1$ in the upper panels and $0.98$ in the lower panels,
in compliance with LHC data. 
The vertical dotted line shows the upper limit on $\lambda_5^i$ from the EDMs and $S$ and $T$ parameters. 
It is evident that, over the $\lambda_5^i$ allowed regions, deviations of the SM-like Higgs couplings to $W^+W^-$ and $ZZ$ pairs induced by CPV 
are negligible, thereby generating no tension against LHC data. 
On the other hand,
the magnitudes of corresponding couplings of the other two neutral Higgs states, $H_2$ and $H_3$,
grow with increasing $\lambda_5^i$. 
Note that $|\xi_V^{H_2}|$ increases rapidly as $s_{\beta-\tilde{\alpha}}$ changes from $1$ to $0.98$, while it does not change considerably with the change in $\tan\beta$.
However, $|\xi_V^{H_3}|$ decreases with growing $\tan\beta$ and with the change of $s_{\beta-\tilde{\alpha}}$ from $1$ to $0.98$.
This is clearly
conducive to establish the $W^+W^-$ and $ZZ$ decays of three Higgs states of the 2HDM Type-I we are considering as a hallmark
signature of CPV.

\begin{figure}[h!]
\begin{center}
\includegraphics[scale=0.26]{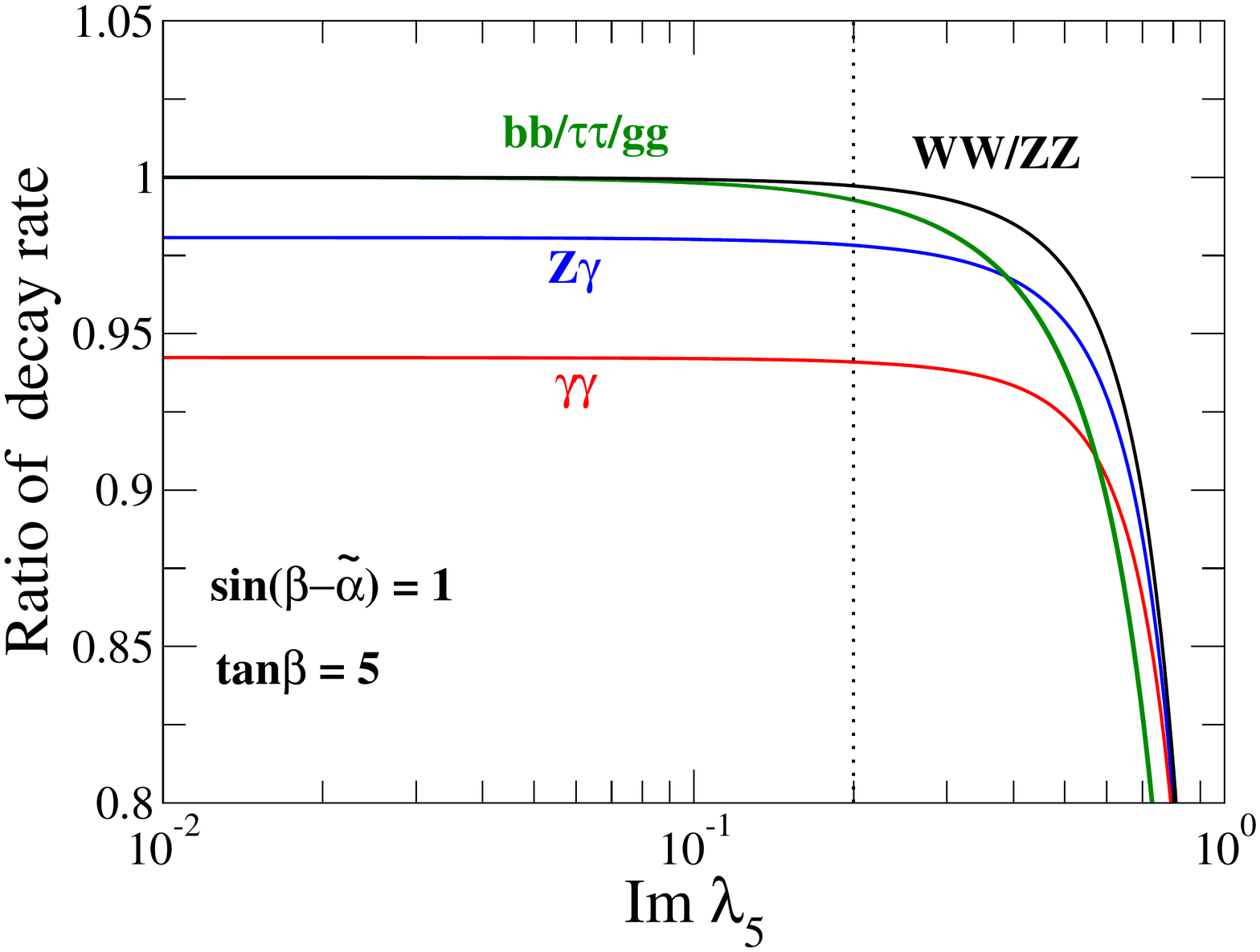} \hspace{5mm}
\includegraphics[scale=0.26]{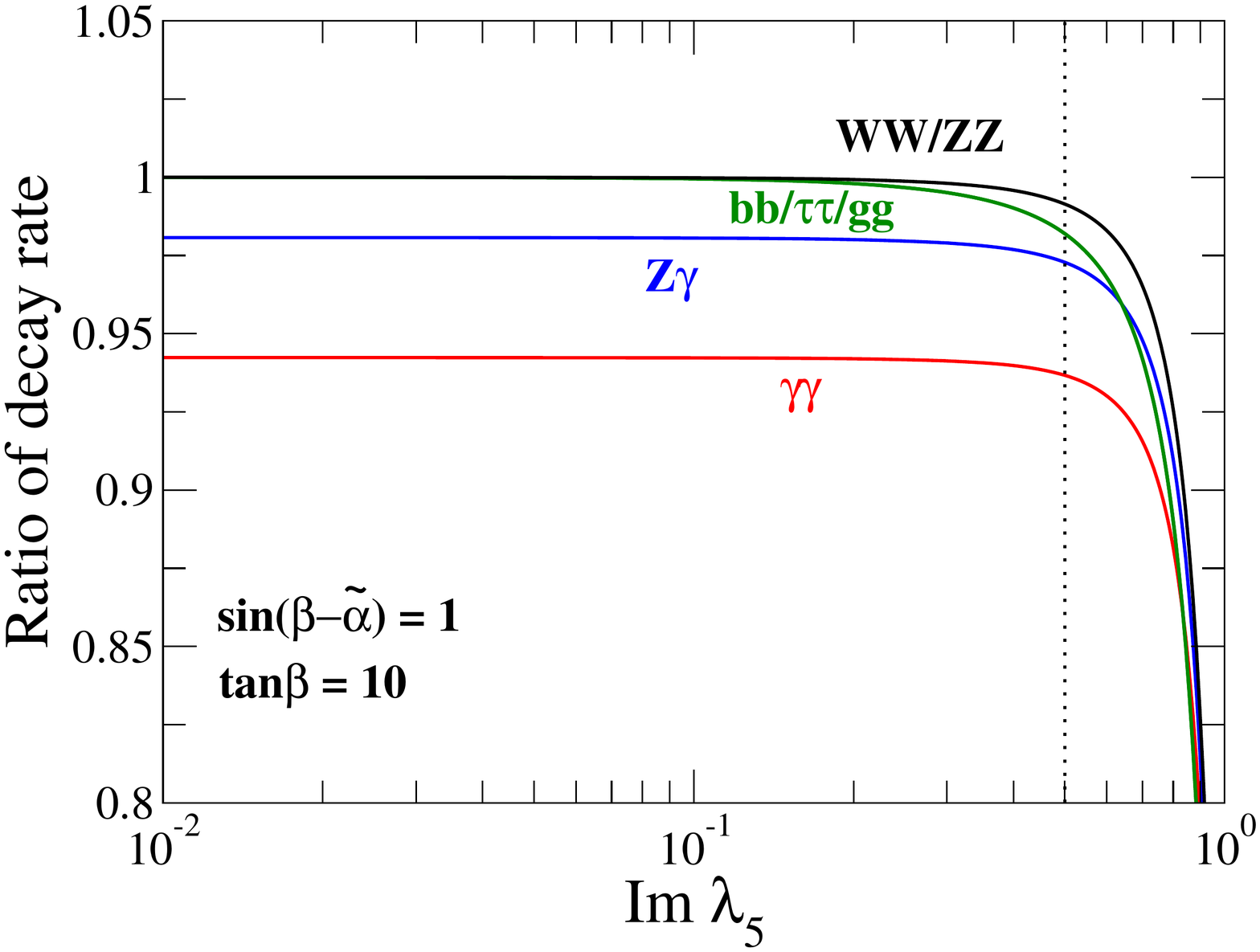} \\
\includegraphics[scale=0.26]{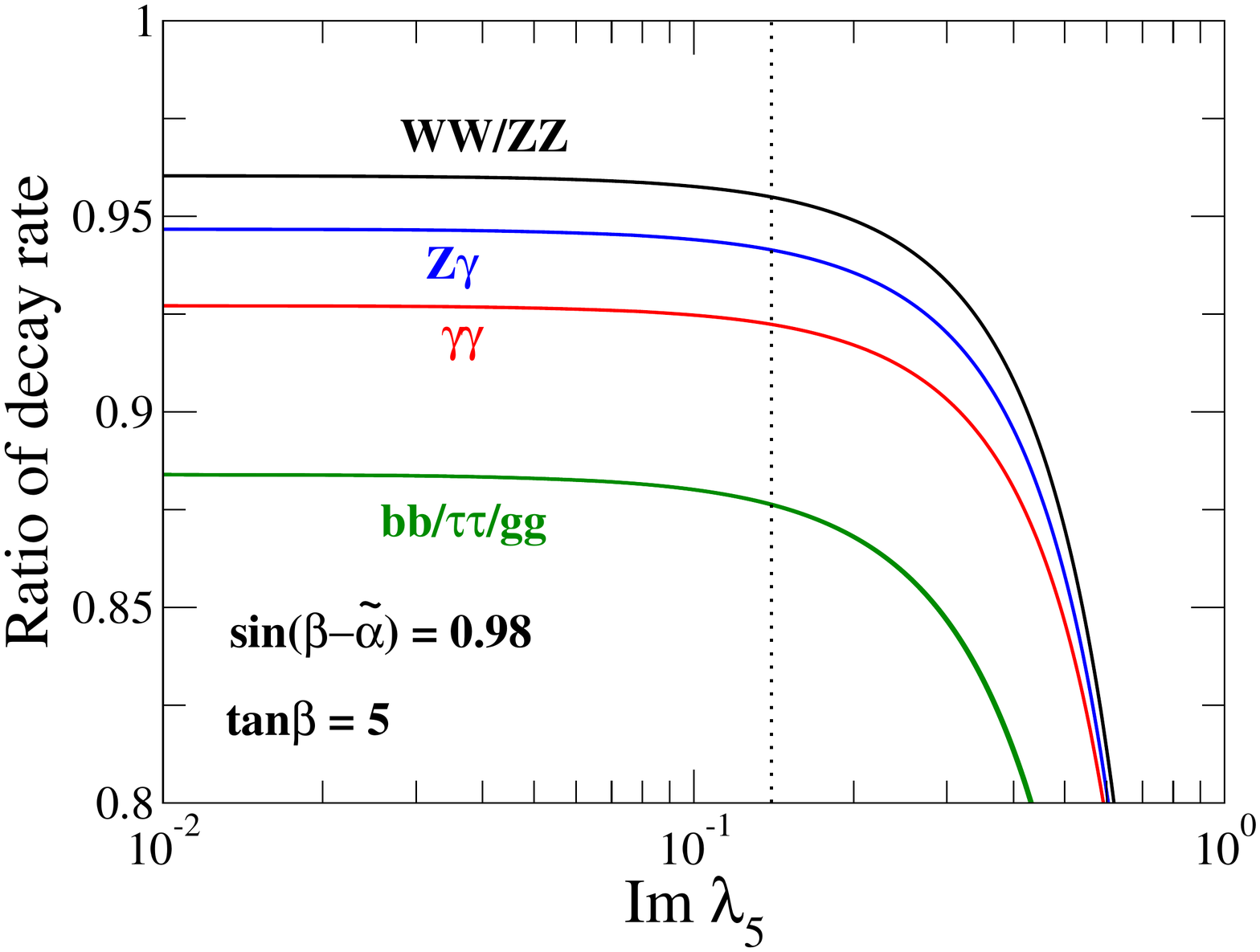}  \hspace{5mm}
\includegraphics[scale=0.26]{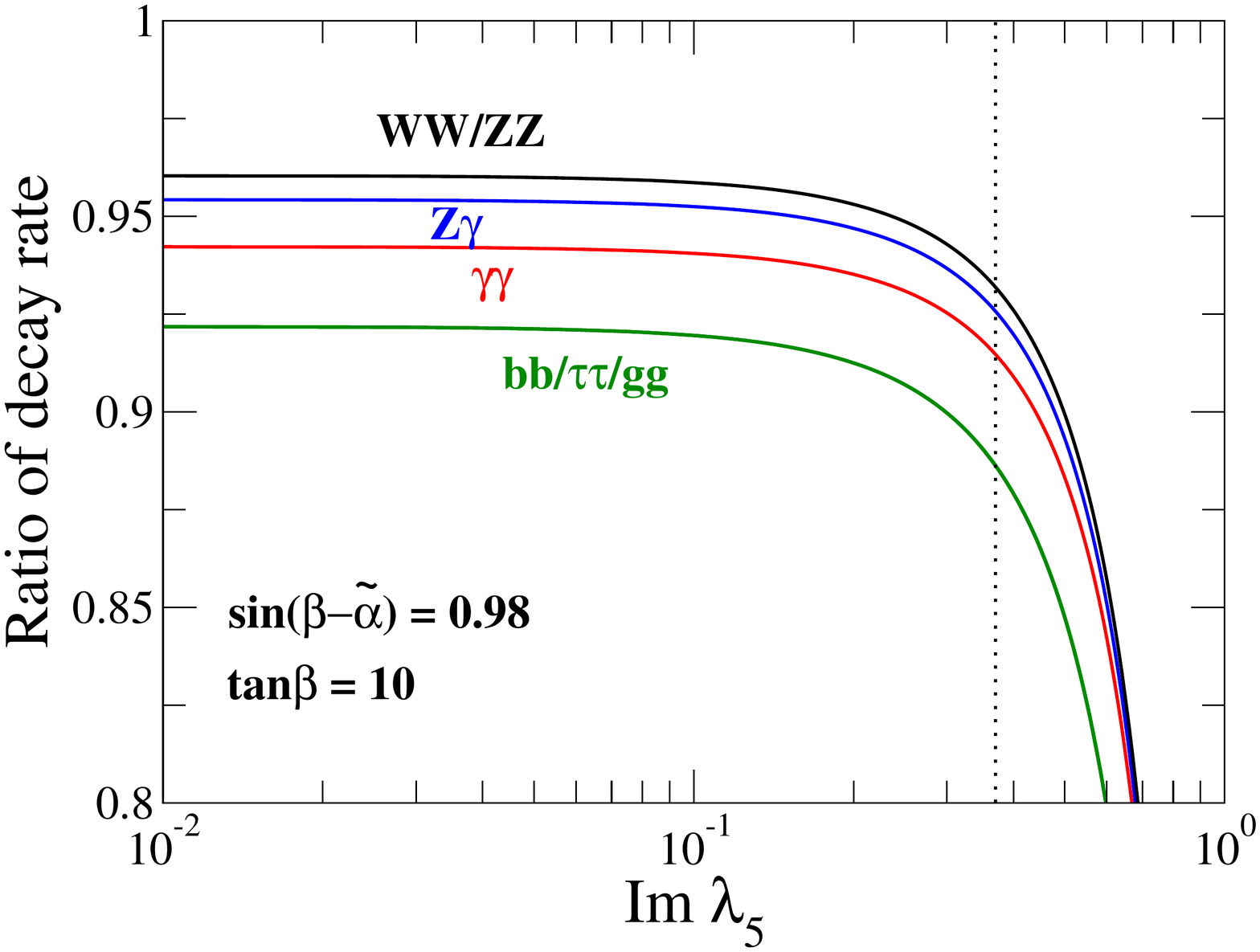} 
\caption{The ratio of decay rates of $h(=H_1)$ to those of the SM Higgs boson $h_{\text{SM}}$ as a function of $\lambda_5^i$ 
for $\tan\beta = 5$ (on the left) and  $\tan\beta = 10$ (on the right). 
The values of $s_{\beta-\tilde{\alpha}}$ are taken to be $1$ and $0.98$ for the upper and lower panels, respectively. 
For all the plots, we take $\tilde{m}_{H}=200$ GeV and $\tilde{m}_A^{}=m_{H^\pm}^{}=250$ GeV.  
The vertical dotted line shows the upper limit on $\lambda_5^i$ from the EDMs and $S$ and $T$ parameters. We take $M=$190 and 180 GeV for the cases of $s_{\beta-\tilde{\alpha}}=1$ and 0.98, respectively. }
\label{ratios}
\end{center}
\end{figure}

In Fig.~\ref{ratios}, we present the ratio of decay rates of  the $H_1$ (identified as the $h$, the SM-like Higgs boson) to those of 
$h_{\text{SM}}$ (the Higgs boson in the SM) 
for two values of $\tan\beta=5$ (on the left) and $\tan\beta=10$ (on the right). 
The vertical dotted line as usual shows the upper limit on $\lambda_5^i$. 
Over the allowed  $\lambda_5^i$ intervals, none of BRs of the SM-like Higgs boson of our
2HDM Type-I deviates significantly from the LHC data, with the possible exception of $b\bar b, \tau^+\tau^-$ and $gg$,
when $s_{\beta-\tilde{\alpha}}$ departs from 1 at small $\tan\beta$. This effect may thus be significant in order
to establish CPV in our scenario in cases where the $H_1$ state is not produced in the SM-like channels presently
investigated and constrained by the LHC, for example, in cascade decays of the heavier Higgs states. We remark 
though that this occurs in a complementary
region of 2HDM Type-I parameter space to the one where treble  $W^+W^-$ and $ZZ$ signals of the neutral Higgs states can be
established, i.e., when  $s_{\beta-\tilde{\alpha}}$  is closer to 1 and $\tan\beta$ is larger.

\begin{figure}[h!]
\begin{center}
\includegraphics[scale=0.26]{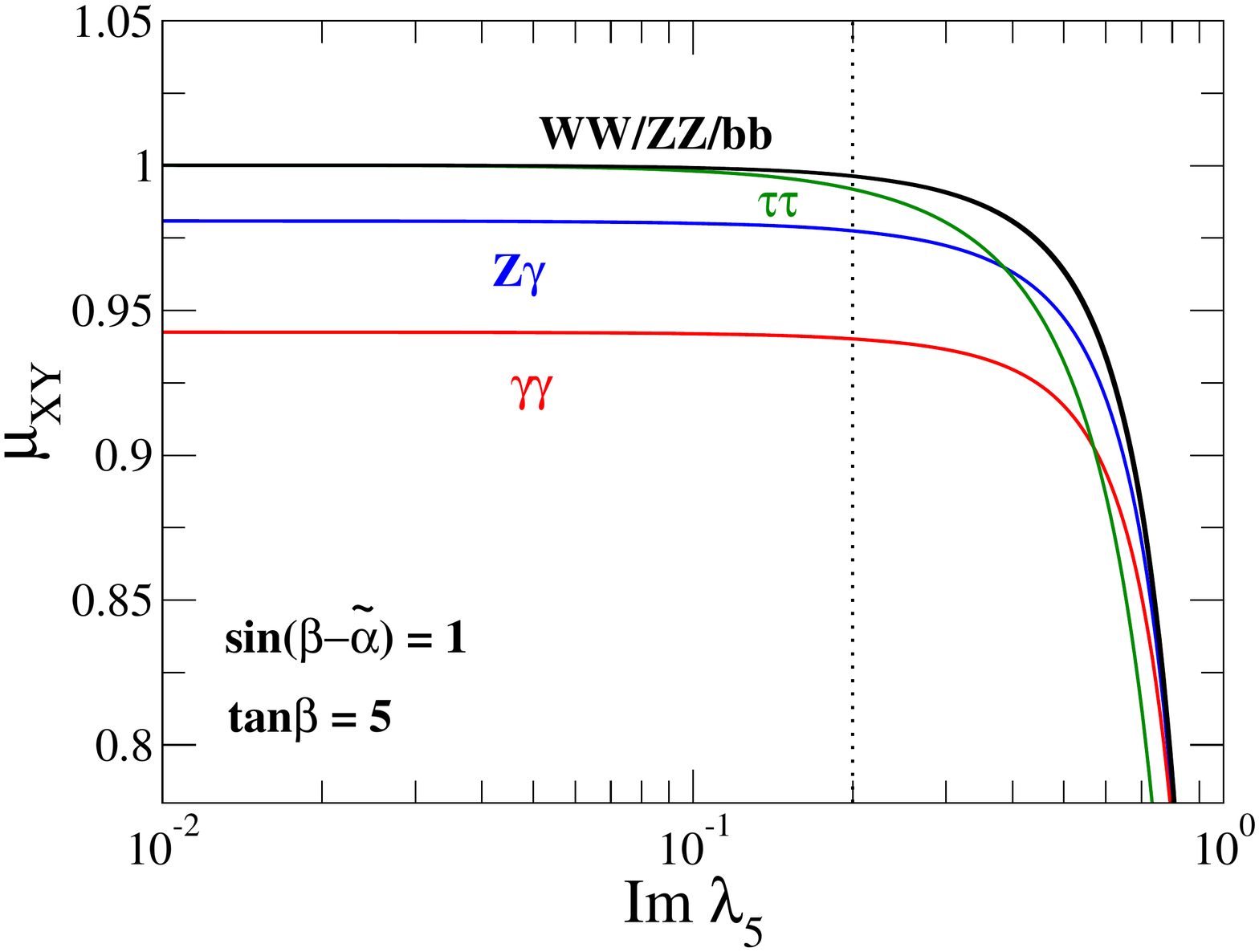} \hspace{5mm}
\includegraphics[scale=0.26]{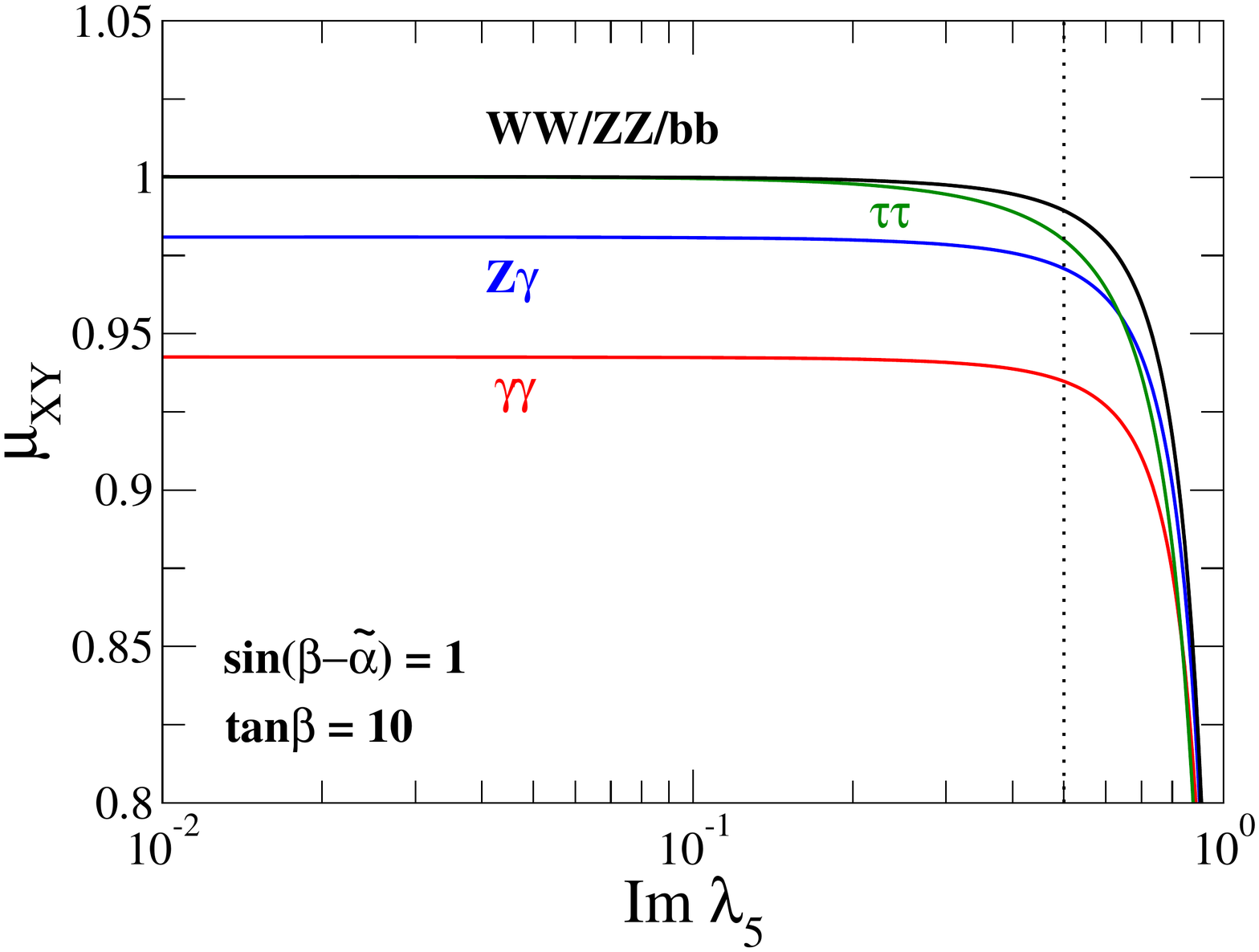} \\
\includegraphics[scale=0.26]{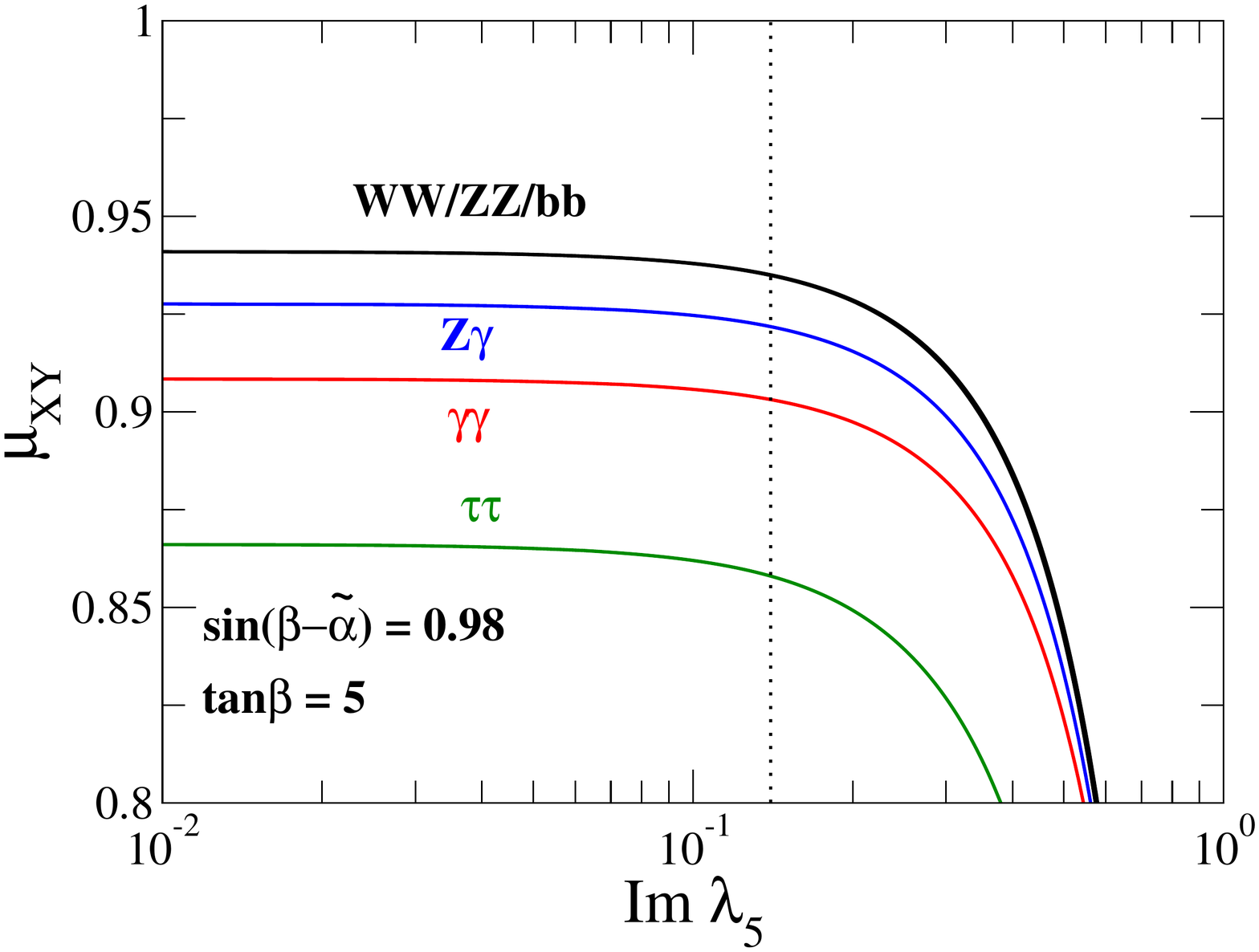}  \hspace{5mm}
\includegraphics[scale=0.26]{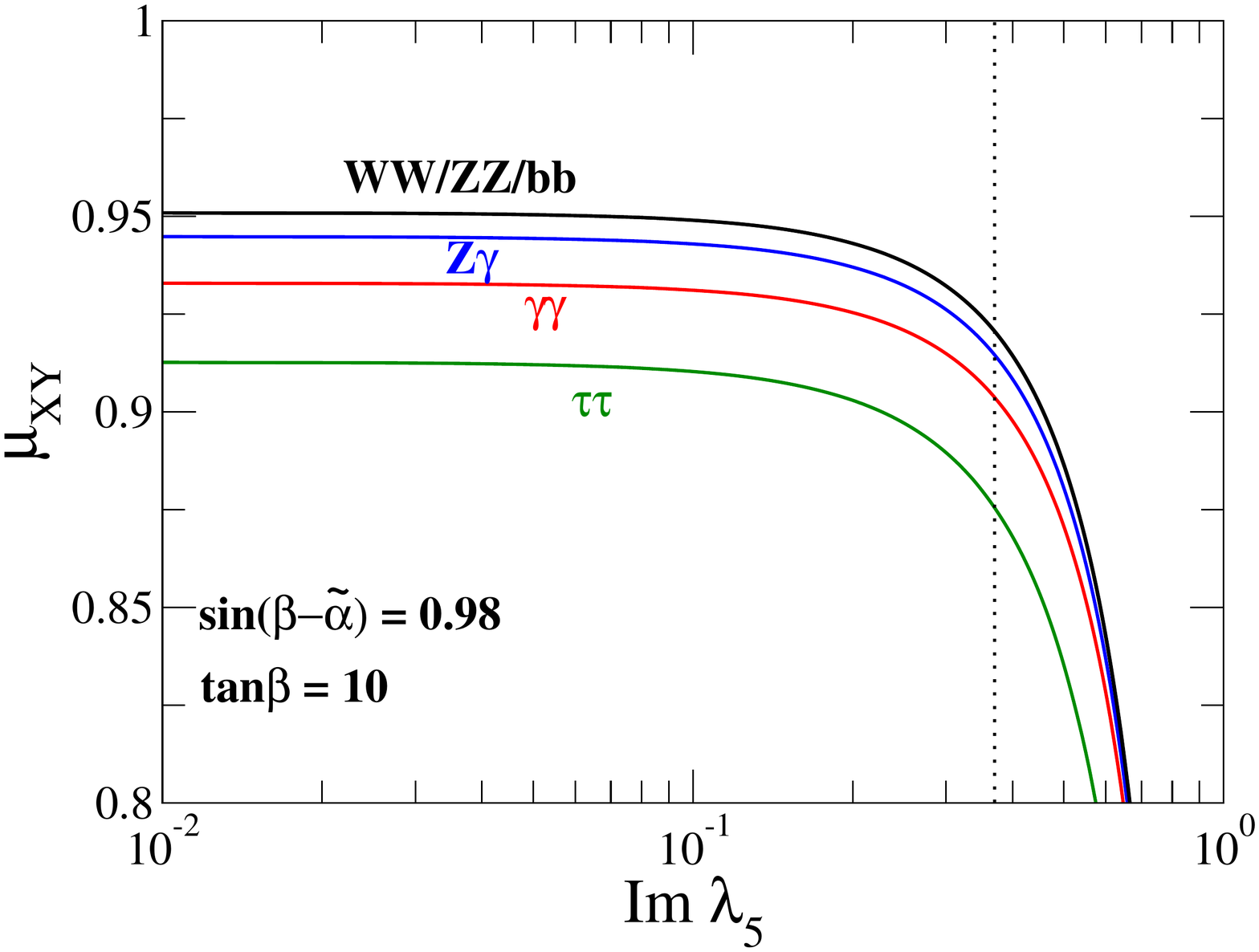} 
\caption{The signal strength for the SM-like Higgs boson $h(=H_1)$ 
as a function of $\lambda_5^i$ for $\tan\beta = 5$ (on the left) and  $\tan\beta = 10$ (on the right). 
The values of $s_{\beta-\tilde{\alpha}}$ are taken to be $1$ and $0.98$ for the upper and lower panels, respectively. 
For all the plots, we take $\tilde{m}_{H}=200$ GeV and $\tilde{m}_A^{}=m_{H^\pm}^{}=250$ GeV.  
The vertical dotted line shows the upper limit on $\lambda_5^i$ from the EDMs and $S$ and $T$ parameters. We take $M=$190 and 180 GeV for the cases of $s_{\beta-\tilde{\alpha}}=1$ and 0.98, respectively. }
\label{signal}
\end{center}
\end{figure}

Fig.~\ref{signal} shows the signal strength, $\mu_{XY}^{}$, of the SM-like Higgs boson $h(=H_1)$, 
defined as
\begin{align}
\mu_{XY}^{}& = \frac{\sigma(gg\to H_1)}{\sigma(gg\to h_{\text{SM}} )}\times \frac{\text{BR}(H_1\to XY)}{\text{BR}(h_{\text{SM}}\to XY)},\quad
XY = W^+W^-,~ZZ,~gg,~\gamma\gamma,~Z\gamma,~\tau^+\tau^-, \\
\mu_{b\bar{b}} &= \frac{\sigma(q\bar{q}\to H_1 V)}{\sigma(q\bar{q}\to h_{\text{SM}} V)}\times \frac{\text{BR}(H_1\to  b\bar{b})}{\text{BR}(h_{\text{SM}} \to b\bar{b})}.
\end{align}
Owing to the interplay between the CPV effects entering directly or indirectly the signal strengths via the production cross
sections, partial decay widths and the total one as seen at the LHC, of the three aforementioned decay modes of the
$H_1$ state, only the $\tau^+\tau^-$ one may carry some evidence of CPV effects, again, for the same
conditions, i.e., when $s_{\beta-\tilde{\alpha}}$ departs from 1 at small $\tan\beta$. Hence, this offers a second
handle to access CPV in the 2HDM Type-I studied here, alternative to the smoking gun signature of the aforementioned
$W^+W^-$ and $ZZ$ decays, as the measurements of the fermionic signal strengths of the SM-like Higgs state will
improve at Run 2 of the LHC.

\begin{figure}[h!]
\begin{center}
\includegraphics[scale=0.26]{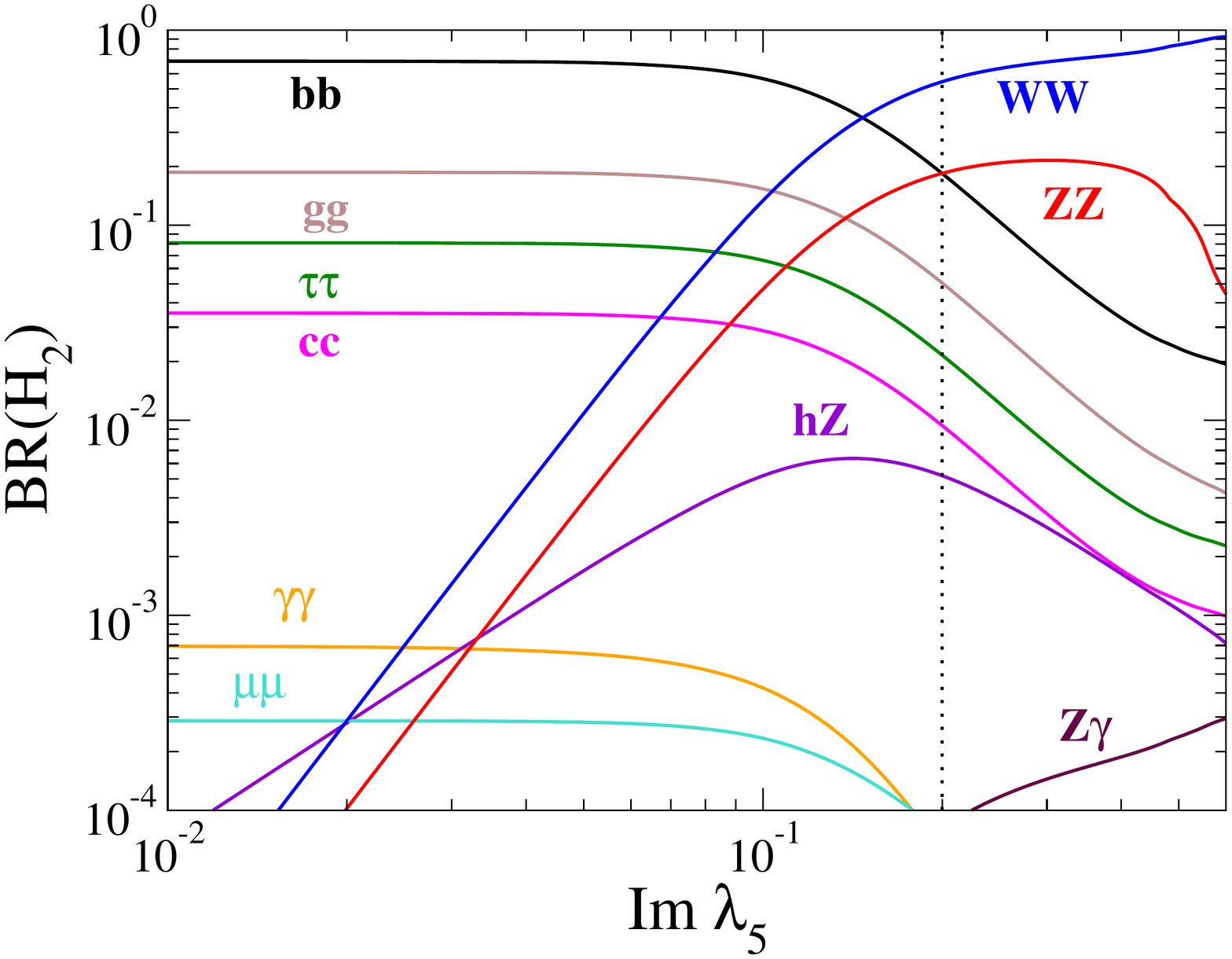} \hspace{5mm}
\includegraphics[scale=0.26]{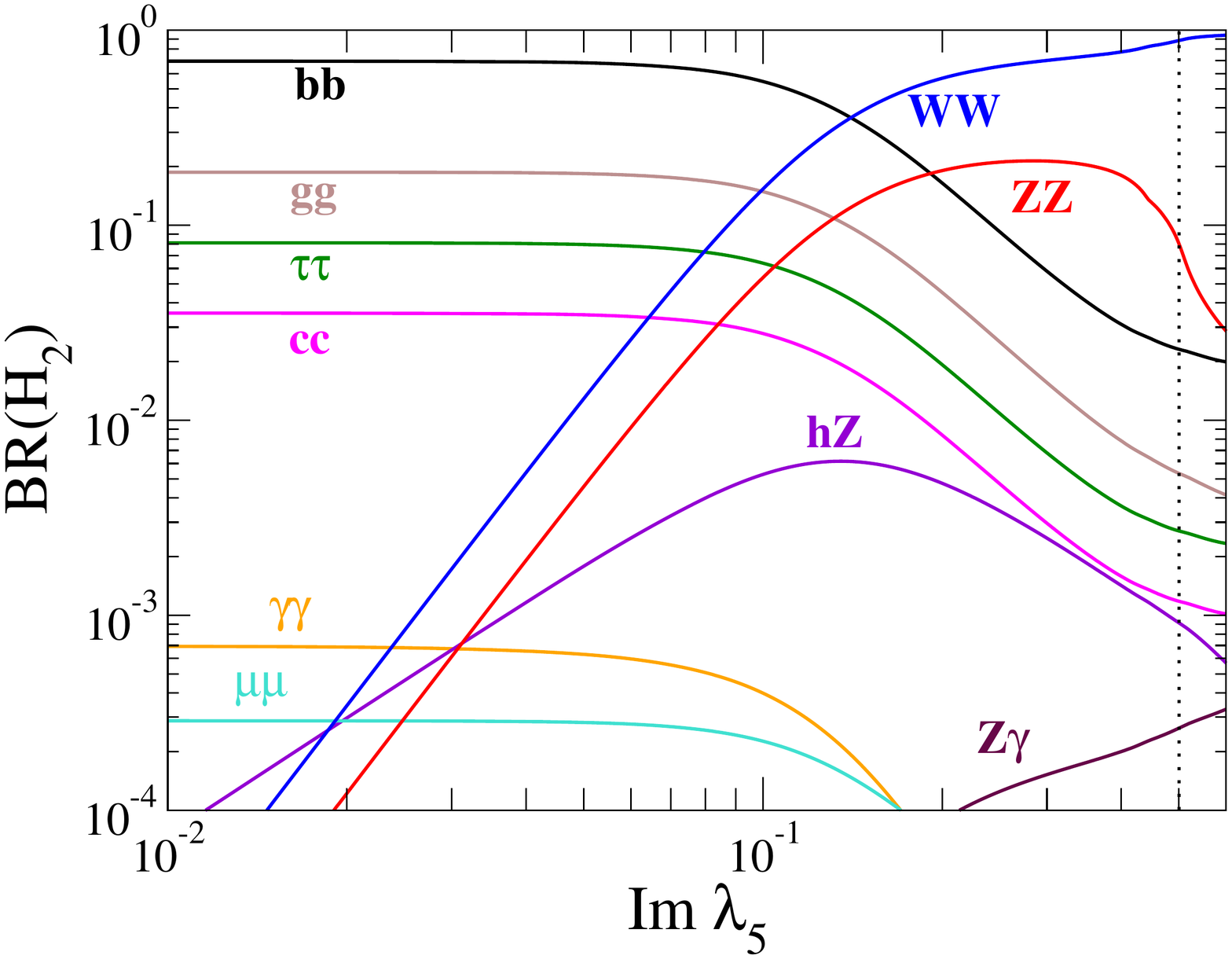} \\
\includegraphics[scale=0.26]{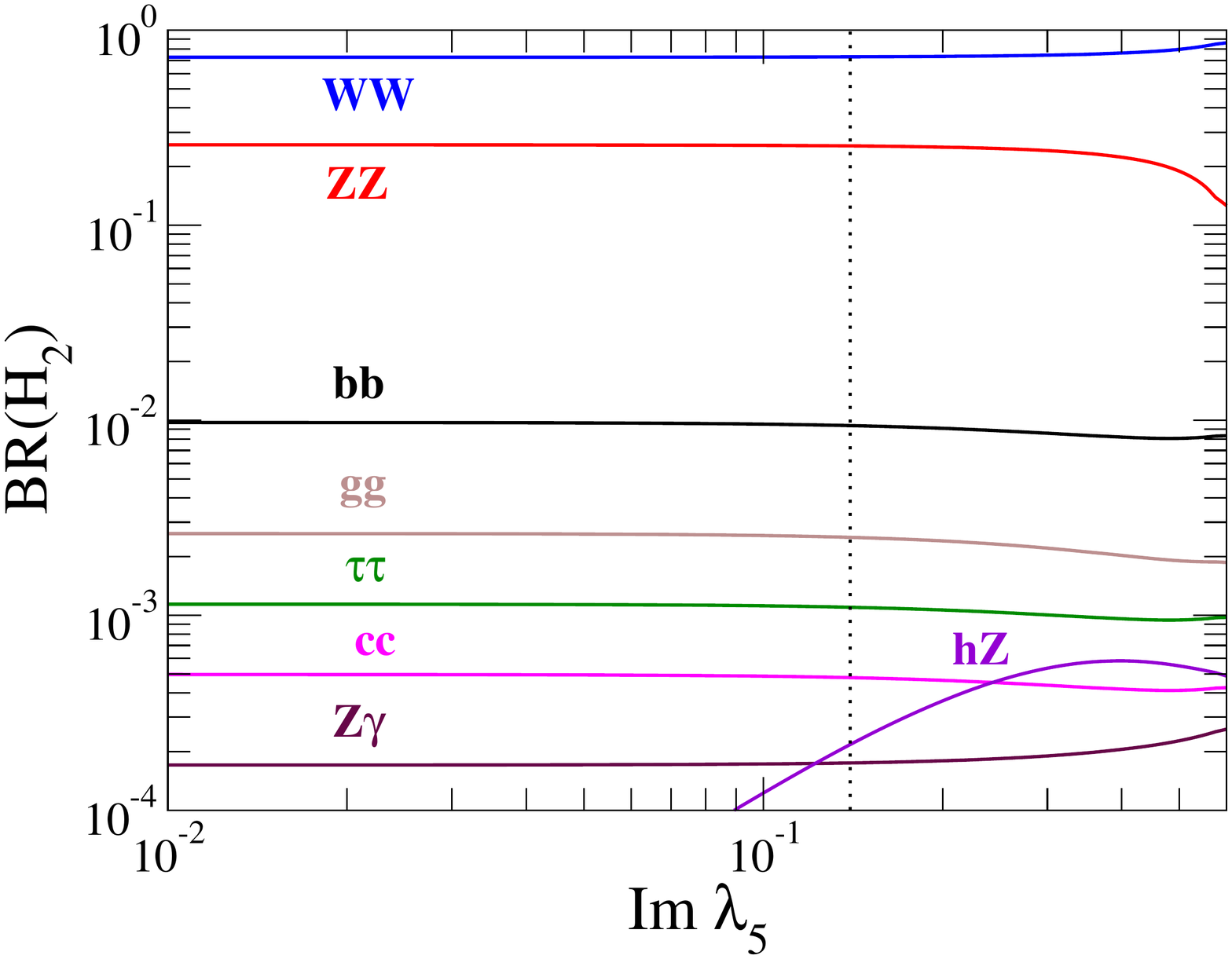}  \hspace{5mm}
\includegraphics[scale=0.26]{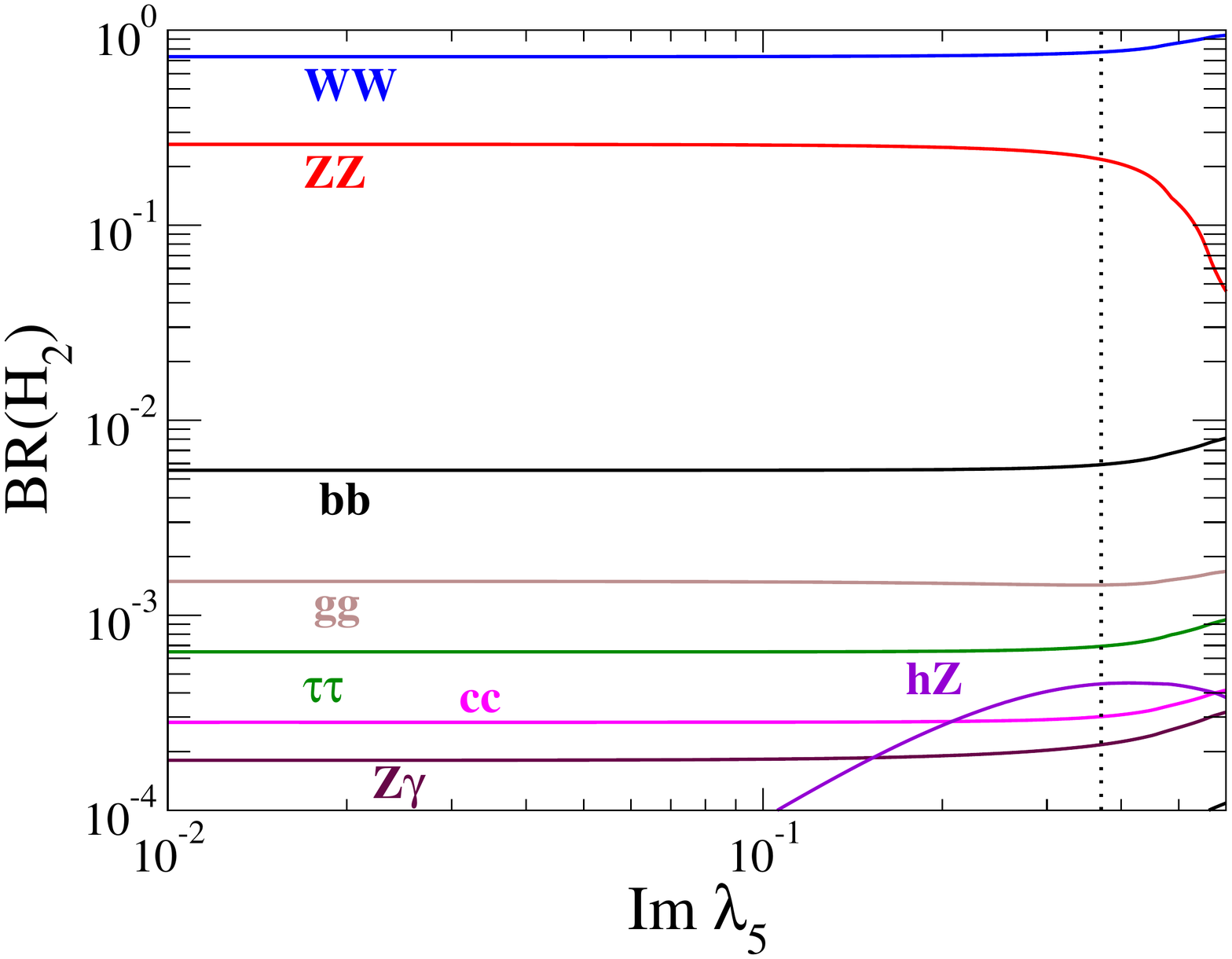} 
\caption{The branching fractions for $H_2$ 
as a function of $\lambda_5^i$ for $\tan\beta = 5$ (on the left) and  $\tan\beta = 10$ (on the right). 
The values of $s_{\beta-\tilde{\alpha}}$ are taken to be $1$ and $0.98$ for the upper and lower panels, respectively. 
For all the plots, we take $\tilde{m}_{H}=200$ GeV and $\tilde{m}_A^{}=m_{H^\pm}^{}=250$ GeV.  
The vertical dotted line shows the upper limit on $\lambda_5^i$ from the EDMs and $S$ and $T$ parameters. We take $M=$190 and 180 GeV for the cases of $s_{\beta-\tilde{\alpha}}=1$ and 0.98, respectively. }
\label{branchingH2}
\end{center}
\end{figure}

\begin{figure}[h!]
\begin{center}
\includegraphics[scale=0.26]{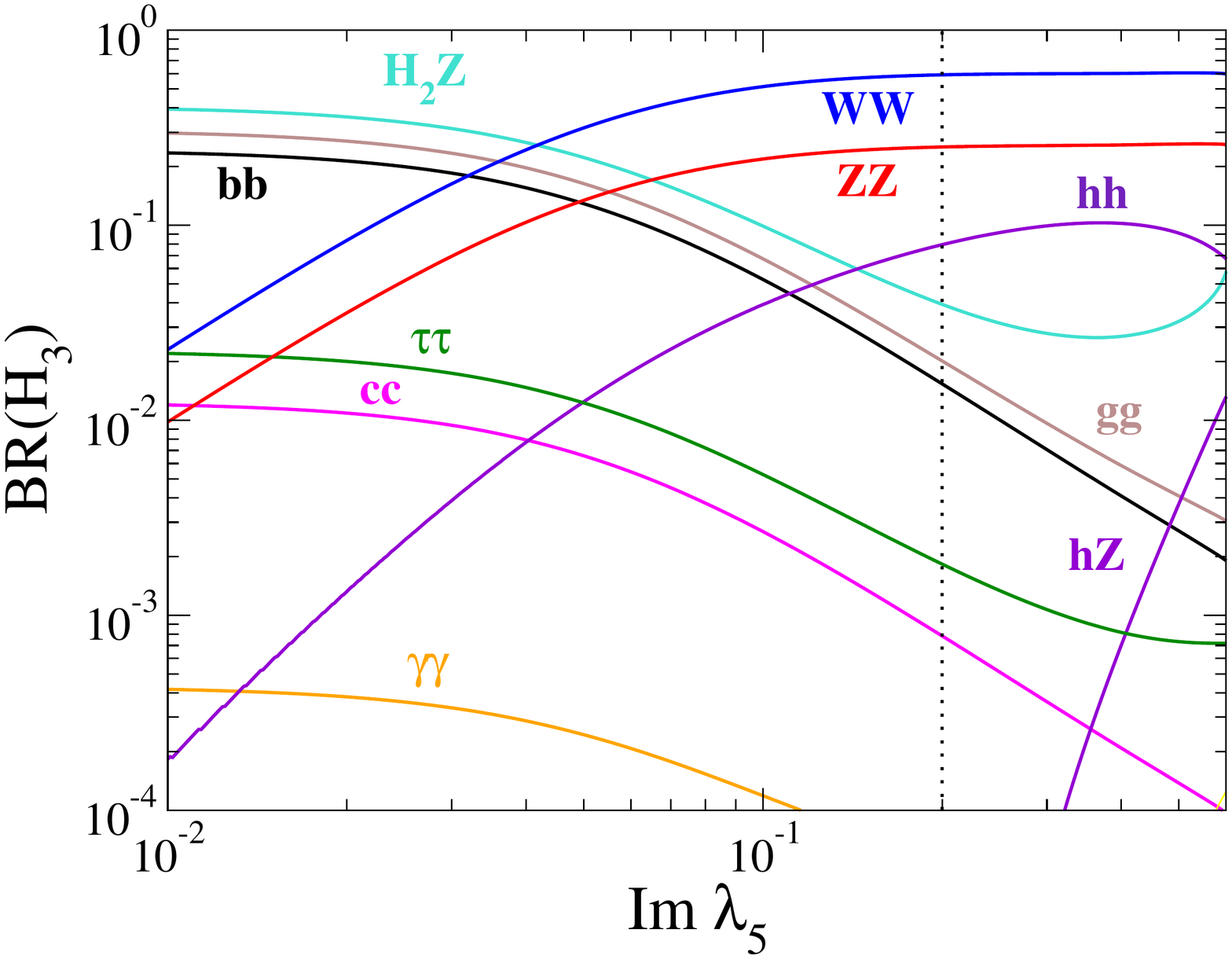} \hspace{5mm}
\includegraphics[scale=0.26]{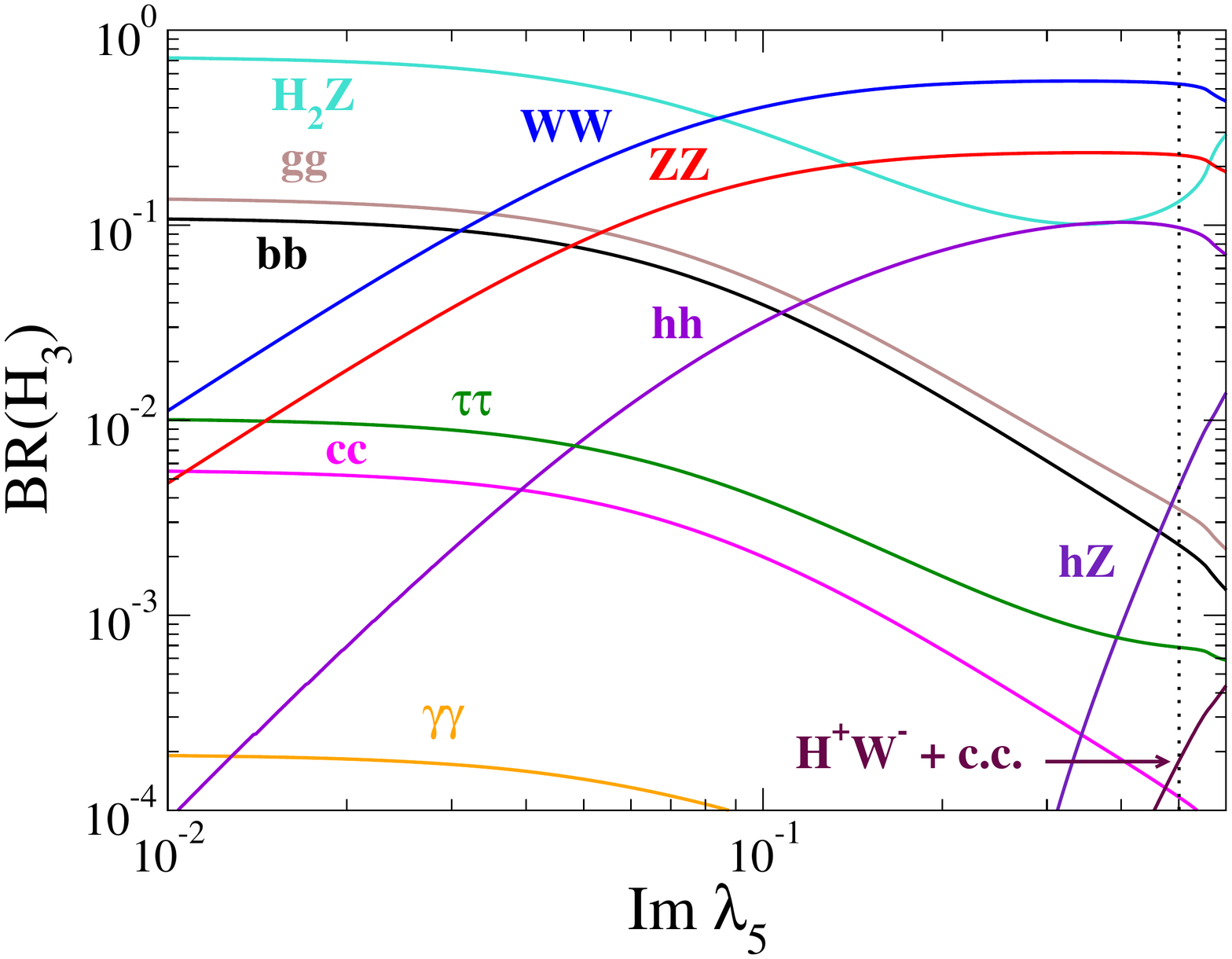} \\
\includegraphics[scale=0.26]{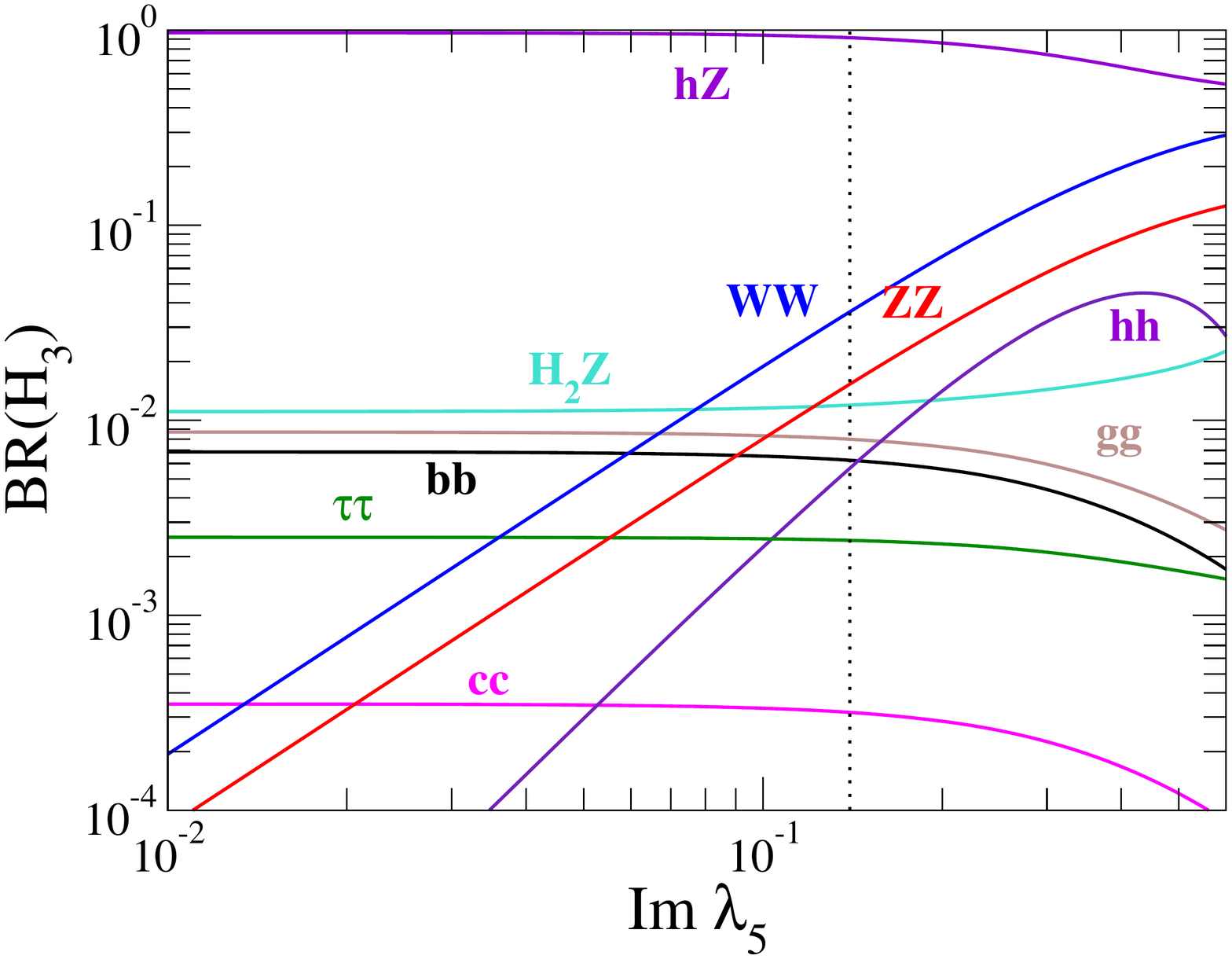}  \hspace{5mm}
\includegraphics[scale=0.26]{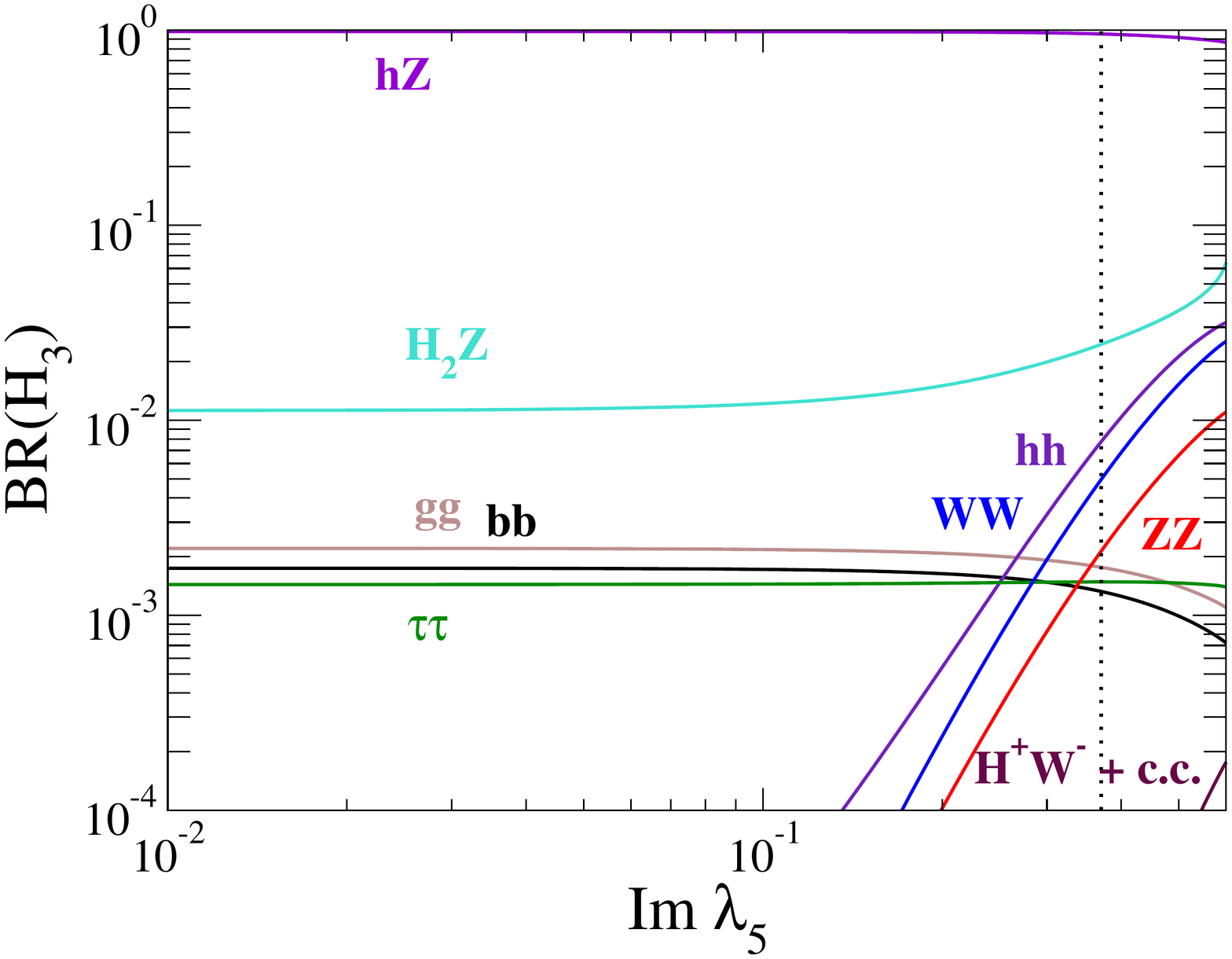} 
\caption{The branching fractions for $H_3$ 
as a function of $\lambda_5^i$ for $\tan\beta = 5$ (on the left) and  $\tan\beta = 10$ (on the right). 
The values of $s_{\beta-\tilde{\alpha}}$ are taken to be $1$ and $0.98$ for the upper and lower panels, respectively. 
For all the plots, we take $\tilde{m}_{H}=200$ GeV and $\tilde{m}_A^{}=m_{H^\pm}^{}=250$ GeV.  
The vertical dotted line shows the upper limit on $\lambda_5^i$ from the EDMs and $S$ and $T$ parameters. We take $M=$190 and 180 GeV for the cases of $s_{\beta-\tilde{\alpha}}=1$ and 0.98, respectively. }
\label{branchingH3}
\end{center}
\end{figure}

\begin{figure}[h!]
\begin{center} 
\includegraphics[scale=0.26]{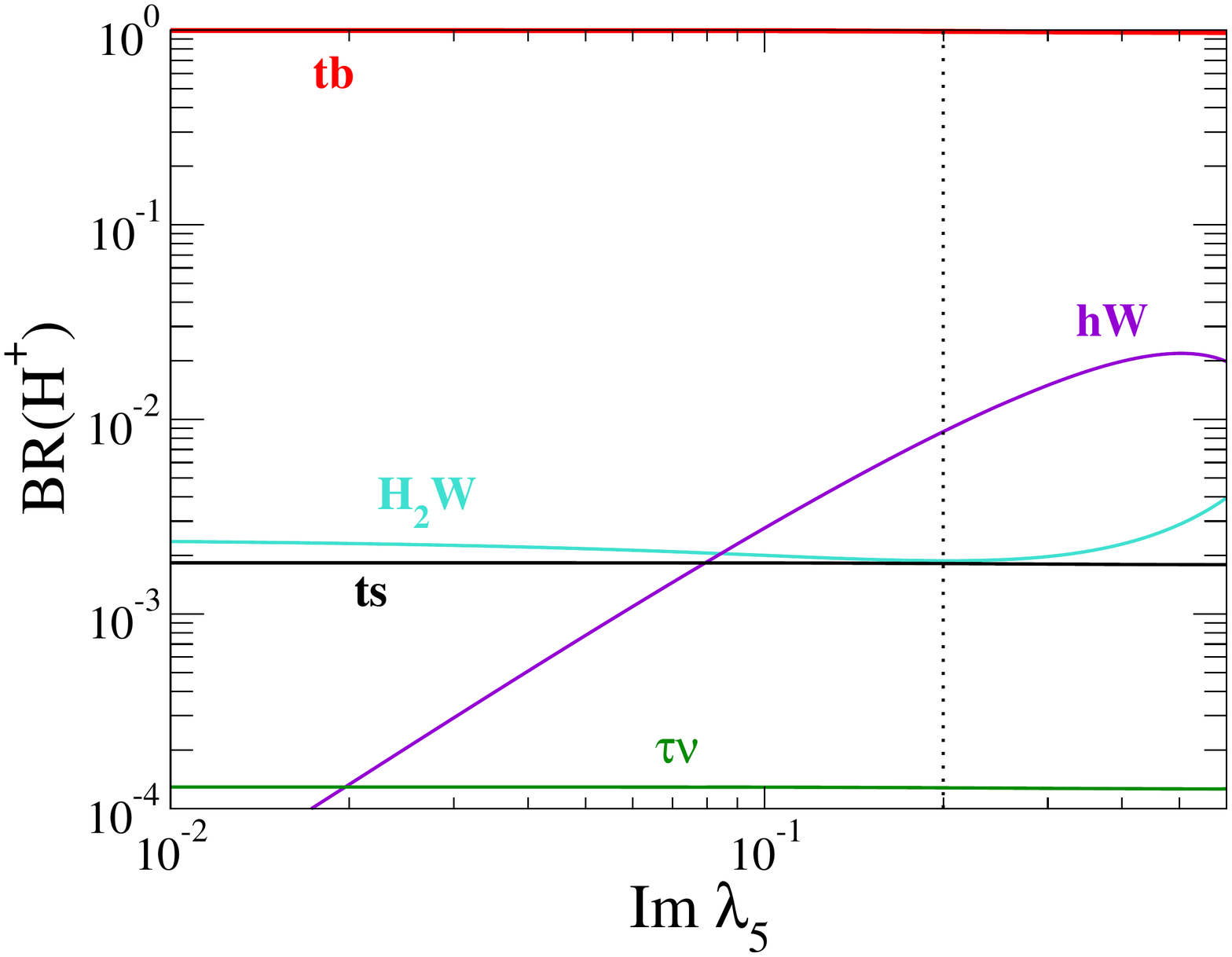} \hspace{5mm}
\includegraphics[scale=0.26]{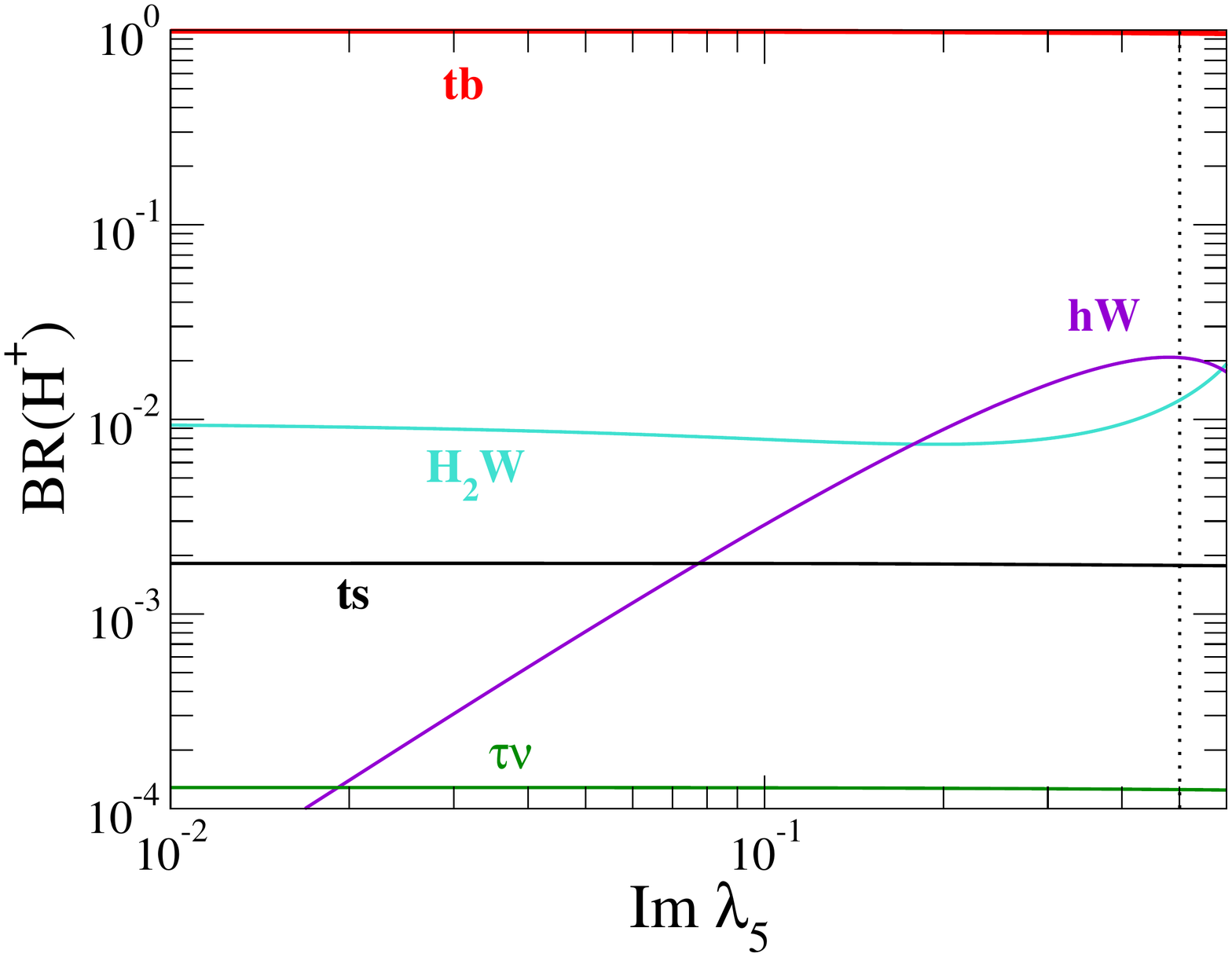} \\
\includegraphics[scale=0.26]{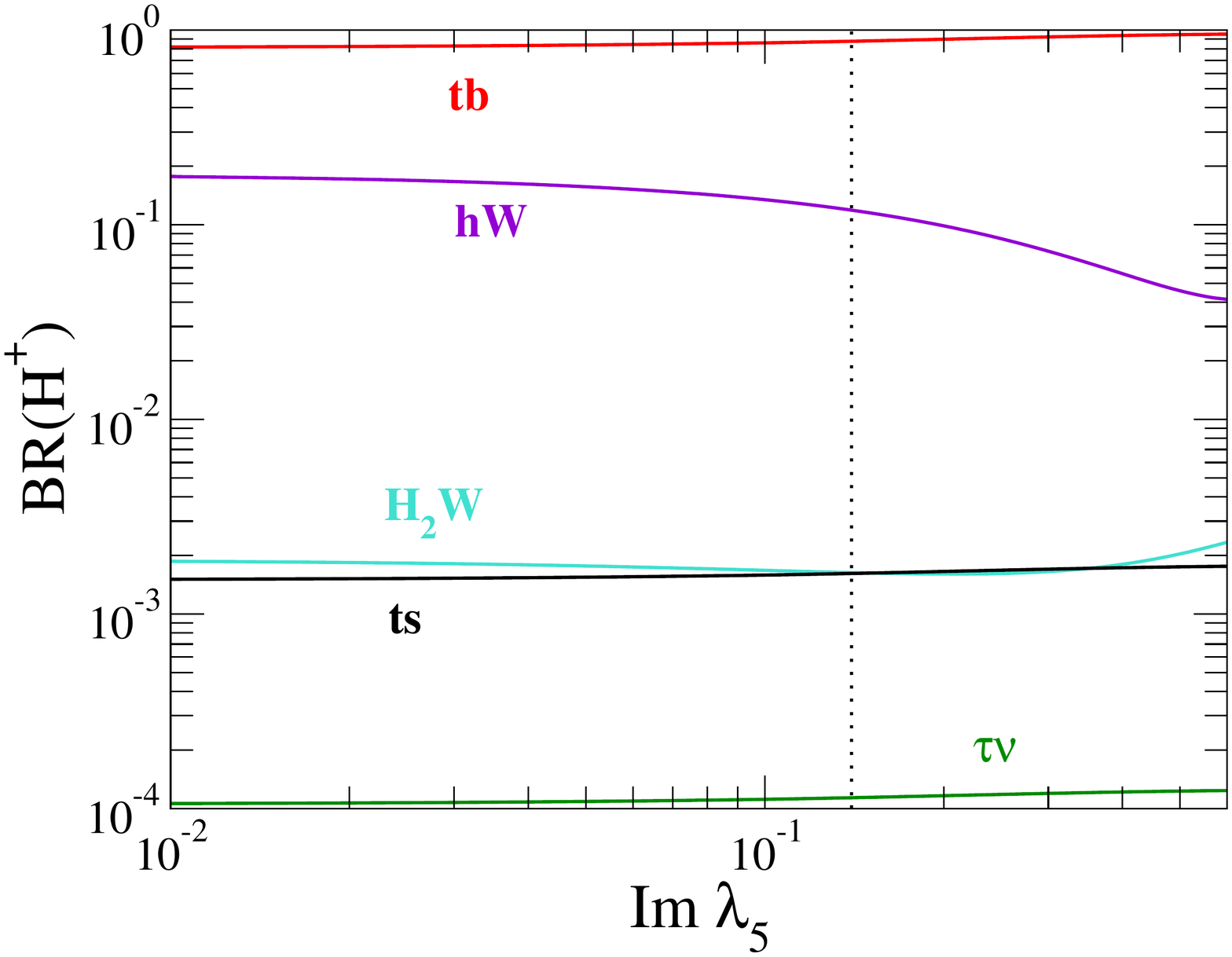}  \hspace{5mm}
\includegraphics[scale=0.26]{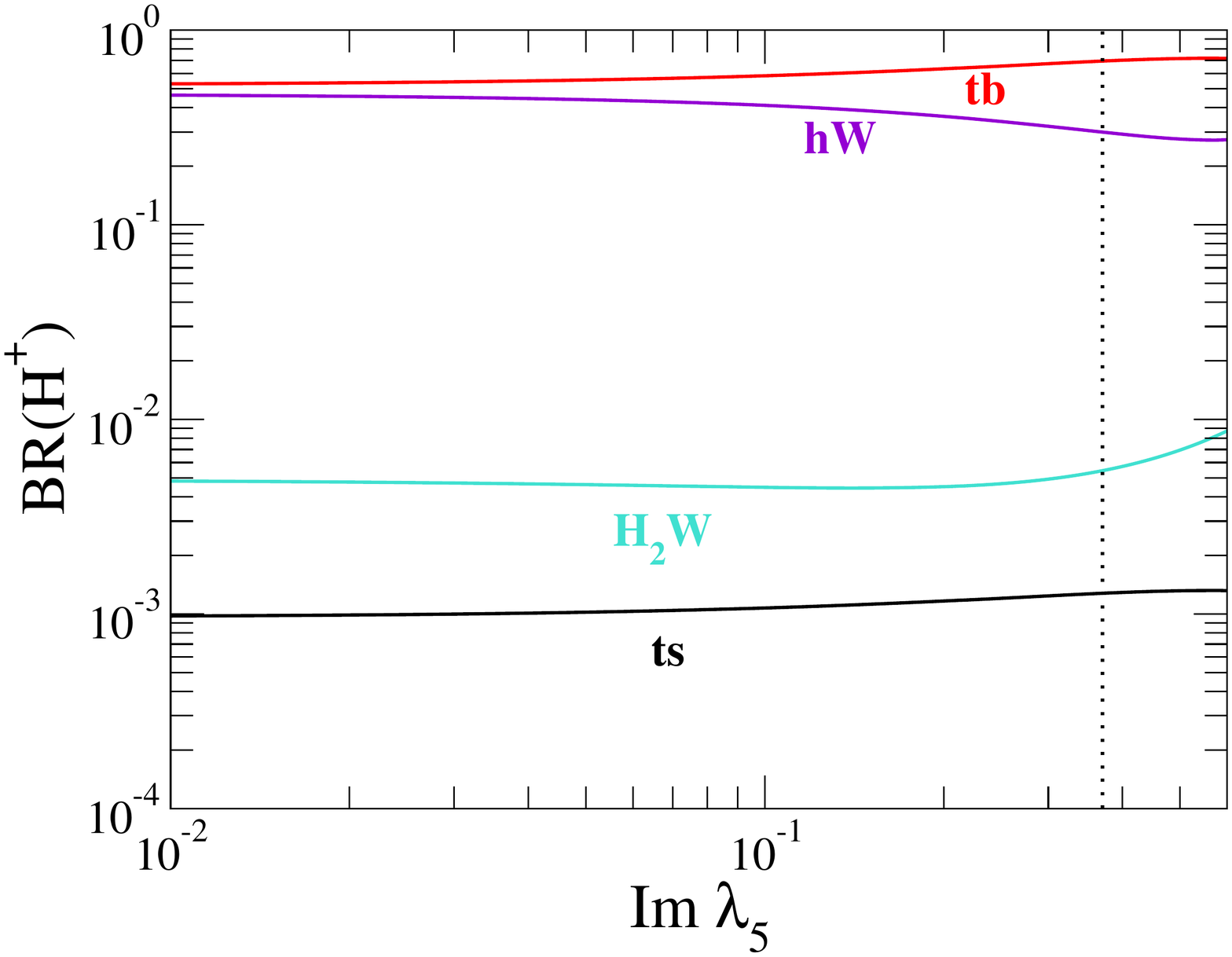} 
\caption{The branching fractions for $H_2$ 
as a function of $\lambda_5^i$ for $\tan\beta = 5$ (on the left) and  $\tan\beta = 10$ (on the right). 
The values of $s_{\beta-\tilde{\alpha}}$ are taken to be $1$ and $0.98$ for the upper and lower panels, respectively. 
For all the plots, we take $\tilde{m}_{H}=200$ GeV and $\tilde{m}_A^{}=m_{H^\pm}^{}=250$ GeV.  
The vertical dotted line shows the upper limit on $\lambda_5^i$ from the EDMs and $S$ and $T$ parameters. We take $M=$190 and 180 GeV for the cases of $s_{\beta-\tilde{\alpha}}=1$ and 0.98, respectively. }
\label{branchingH+}
\end{center}
\end{figure}

Fig.~\ref{branchingH2} shows the BRs of the second lightest neutral Higgs boson, $H_2$, 
as a function of $\lambda_5^i$ for $\tan\beta = 5$ (on the left) and  $\tan\beta = 10$ (on the right).
We take $s_{\beta-\tilde{\alpha}}=1$ (upper panels) and 0.98 (lower panels). Similarly,
Fig.~\ref{branchingH3}  does so for  the heaviest neutral Higgs boson, $H_3$. By contrasting the two,
it is evident that the largest $W^+W^-$ and $ZZ$ rates are {\sl simultaneously} found, as intimated, for large 
$\tan\beta$ and $H_1$ couplings very SM-like. 
%
Note that $H_1,H_2,H_3 \to WW/ZZ$ are all large simultaneously only in the upper top plot of Figs.~\ref{branchingH2}-\ref{branchingH3} already well below the EDM limit, whereas in the other 3 plots this decay rate can be large only very close to the EDM limit (in the top left plot, $H_2 \to WW/ZZ$ becomes dominant essentially where the parameter space is starting to be ruled out) or else only 2 of the channels can be large at the same (in the bottom plots, $H_3 \to WW/ZZ$ is always subleading).
Another possible hallmark signal of CPV could be the $hZ$ one, having assessed  that
current experimental constraints force the $H_1\equiv h$ state of the 2HDM Type-I to be essentially CP-even. 
Under this condition, in fact, to establish $hZ$, it would mean for both $H_2$ and $H_3$ to have a CP-odd nature,
hence unlike the case of the corresponding CPC version of our scenario. Unfortunately, the $H_2$ and $H_3$ BRs
are never large simultaneously in the allowed $\lambda_5^i$ regions. As for
other decay modes, while interesting patterns emerge, we notice that none of these can be taken as a direct evidence
of CPV as they all exist already in the CPC case for both the heavy Higgs states.

Fig.~\ref{branchingH+} shows the BRs  of the charged Higgs bosons, $H^\pm$, 
as a function of $\lambda_5^i$ for $\tan\beta = 5$ (on the left) and  $\tan\beta = 10$ (on the right). As usual,
we take $s_{\beta-\tilde{\alpha}}=1$ (upper panels) and 0.98 (lower panels). As just remarked for most of the
$H_2$ and $H_3$ decay rates, here, again, interesting decay patterns emerge, yet all the  possible final states already exist in the CPC case of the 2HDM Type-I. This also includes the case of $hW^\pm$ and $H_2W^\pm$ decays (in the CPC 2HDM Type-I the latter would be either $HW^\pm$ or $AW^\pm$), which show an interesting interplay (as function
of $\lambda_5^i$) generally unseen in the CPC case, which may eventually help
as confirmation of CPV being present  in the charged Higgs sector too.   

\begin{center}
\begin{table}[t]
\begin{scriptsize}
\begin{tabular}{|p{10mm}||p{16mm}|p{16mm}|p{18mm}|p{16mm}|p{17mm}|p{17mm}|p{18mm}|}
\hline & $\sigma(gg\to H_2)$ & $\sigma(gg\to H_3)$ & $\sigma(gb\to H^\pm t)$ & $pp \to H_2 H_3$ & $pp \to H_2 H^\pm$  &
$pp \to H_3 H^\pm$ & $pp \to H^+ H^-$\\[2mm]  \hline
$t_\beta=5$  &0.79(0.90)  &4.22(4.83)  &0.057(0.070) &9.0(10)$\times 10^{-3}$ &18(21)$\times 10^{-3}$ &12(14)$\times 10^{-3}$ &6.9(7.9)$\times 10^{-3}$   \\[2mm]\hline
$t_\beta=10$ &0.20(0.23)  &1.06(1.22)  &0.014(0.018) &8.9(10)$\times 10^{-3}$ &18(21)$\times 10^{-3}$ &12(14)$\times 10^{-3}$ &6.9(7.9)$\times 10^{-3}$  \\[2mm]  \hline
\end{tabular}
\end{scriptsize} 
\caption{Production cross sections (in the unit of pb) for extra Higgs bosons at the LHC with the collision energy of 13 (14) TeV in the case of $\tan\beta=5$ and 10. 
We take $\lambda_5^i=0.1$, $\tilde{m}_{H}=200$ GeV, $m_{H^\pm}=\tilde{m}_{A}=250$ GeV and $s_{\beta-\tilde{\alpha}}=1$. 
} \label{CS}
\end{table}
\end{center}
Clearly, in order so see the smoking gun signals described above, one should make sure that $H_2$, $H_3$ and $H^\pm$ states
of the 2HDM Type-I can be copiously produced at the LHC. Hence, we 
finally calculate their production cross sections at the LHC. 
For the neutral Higgs bosons, there are two dominant production processes, namely, 
the gluon fusion process $gg \to H_2,~H_3$ and the pair production $pp \to Z^* \to H_2H_3$. 
For the $H^\pm$ case, there are the $gb$ fusion process $gb \to H^\pm t$ and the pair production 
$pp \to \gamma^*/Z^* \to H^+H^-$. 
In addition to these processes, there are are also mixed modes, i.e., where neutral and charge Higgs states are produced
together via
$pp \to W^* \to H^\pm H_2$ and $pp \to W^* \to H^\pm H_3$. 

The cross section of the gluon fusion process is calculated by 
\begin{align}
&\sigma(gg \to H_2) = \left.\sigma(gg \to h_{\text{SM}})\right|_{m_{h_{\text{SM}}}=m_{H_2}^{}}\times \frac{\Gamma(H_2\to gg)}{\Gamma(h_{\text{SM}}\to gg )}, \\
&\sigma(gg \to H_3) = \left.\sigma(gg \to h_{\text{SM}})\right|_{m_{h_{\text{SM}}}=m_{H_3}^{}}\times \frac{\Gamma(H_3\to gg)}{\Gamma(h_{\text{SM}}\to gg )},  
\end{align}
where $\sigma(gg \to h_{\text{SM}})$ and $\Gamma(h_{\text{SM}}\to gg )$ are 
the gluon fusion cross section and the decay rate of $h_{\text{SM}}\to gg$ for the SM Higgs boson $h_{\text{SM}}$, respectively. 
From Ref.~\cite{HiggsXS}, $\sigma(gg \to h_{\text{SM}})$ is given to be 18.35 pb (21.02 pb) with the collision energy of 13 (14) TeV. 
For the other processes, we calculate these cross sections ourselves. 
The results are listed in Tab.~\ref{CS} with the collision energy of 13 (14) TeV using {\tt CTEQ6L} \cite{Pumplin:2002vw} as 
Parton Distribution Functions (PDFs) at the scale $\mu=\hat s$.  We notice that all cross sections are
in the ${\cal O}(10)$--${\cal O}(1000)$ range, so that the 2HDM Type-I scenario with CPV discussed here would most
likely be probed fully in the years to come, if not at the standard LHC already, certainly at the  tenfold
luminosity increase foreseen at the Super-LHC \cite{Gianotti:2002xx}.

\section{Discussion and conclusion}

\label{conclusion}

In this paper we have studied CPV 2HDMs with a softly-broken $Z_2$ symmetry which is imposed to 
avoid dangerous FCNCs. We have analysed in detail the constraints (mainly from the EDMs and $S,T$ parameters)
and LHC predictions in the Type-I 2HDM in particular.

We have first highlighted possible CPV effects onto the lightest Higgs state of this scenario, $H_1$. Herein, 
deviations from the SM-like behaviour induced by CPV in our scenario, being small
and indirect, while possibly measurable (in fermionic decays) and interesting per se, may be difficult to interpret as such.
In fact, 
the gold plated smoking gun signature of the CPV 2HDM  Type-I is the decay of  both $H_2$ and $H_3$ into weak gauge boson pairs.
Experimentally this will require the observation of all three neutral Higgs bosons $H_{1,2,3}$
decaying into $W^+W^-$ and/or $ZZ$ states. In order to resolve the two heavy neutral Higgs bosons, $H_{2,3}$,
they must be sufficiently non-degenerate with a mass splitting greater than say 10 GeV, which we have seen to be
realisable in our scenario.
For example, for one of the benchmarks considered here, we have $m_{H^\pm}\approx m_{H_3} \approx 250$ GeV and $m_{H_2} \approx 200$ GeV, with a mass splitting of about 50 GeV. Further confirmation of the mixed CP-nature of
the heavy neutral Higgs states could come from their $hZ$ decays, in presence of a light Higgs state
which is essentially SM-like in its quantum numbers, $H_1\equiv h_{\rm SM}$. As for the charged Higgs sector,
indirect evidence of CPV induced by the neutral Higgs states could be seen in the interplay between $H^\pm \to hW^\pm$ and $H_2W^\pm$ decays.

The production cross sections of all heavy states $H_2$, $H_3$ and $H^\pm$ must also be sufficiently large, which we have shown to possibly be the case if both the
standard and high luminosity conditions of the LHC are considered. 

In summary, the 2HDM Type-I is a framework which can implement explicit CPV effects at tree level, free from both theoretical flaws and experimental constraints, that can be probed at the LHC. 

\section*{Acknowledgements}

SFK acknowledges partial support from the STFC Consolidated Grant ST/J000396/ 1 and European Union FP7 ITN-INVISIBLES (Marie Curie Actions, PITN-GA-2011-289442). 
SM is financed in part through the NExT Institute and from the STFC Consolidated Grant ST/ J000396/1. He also acknowledges the H2020-MSCA-RICE-2014 grant no. 645722 (NonMinimalHiggs).
VK's research is financially supported by the Academy of Finland project ``The Higgs Boson and the Cosmos'' and  project 267842.
KY is supported by a JSPS postdoctoral fellowships for research abroad.

\appendix

\section{Higgs trilinear couplings}

The trilinear neutral Higgs boson couplings $\lambda_{ijk}$ defined in Eq.~(\ref{lamijk}) are given by 
\begin{align}
&\lambda_{333} = \lambda_{223} = -\frac{1}{3}\lambda_{113} = \frac{v}{4}\lambda_5^i \sin2\beta, \\
&\lambda_{123} = -v\lambda_5^i \cos2\beta, \\
&\lambda_{222} = \lambda_{233} = \frac{v}{8}\left[\lambda_2-\lambda_1 + (\lambda_1+\lambda_2 -2\lambda_{345})\cos2\beta \right]\sin2\beta, \\
&\lambda_{112} = -\frac{3v}{8}\left[\lambda_1-\lambda_2+ (\lambda_1 + \lambda_2 -2\lambda_{345})\cos2\beta \right]\sin2\beta, \\
&\lambda_{111} = \frac{v}{16}\left[3(\lambda_1+\lambda_2) + 2\lambda_{345} + 4(\lambda_1-\lambda_2)\cos2\beta
+ (\lambda_1+\lambda_2 -2\lambda_{345})\cos4\beta\right], \\
&\lambda_{122} = \frac{v}{16}\left[3(\lambda_1+\lambda_2) + 2\lambda_{345} -3(\lambda_1+\lambda_2 -2\lambda_{345})\cos4\beta\right], \\
&\lambda_{133} = \frac{v}{16}\left[\lambda_1+\lambda_2 +16(\lambda_3+\lambda_4)-10\lambda_{345} -(\lambda_1+\lambda_2 -2\lambda_{345})\cos4\beta\right]. 
\end{align}
The $h_1'H^+H^-$ and $h_2'H^+H^-$ couplings are given by
\begin{align}
\lambda_{h_1' H^+ H^-} &= \frac{v}{8}\left[\lambda_1 + \lambda_2 +8\lambda_3 -2\lambda_{345}-(\lambda_1+\lambda_2 -2\lambda_{345})c_{4\beta} \right] , \\
\lambda_{h_2' H^+ H^-} &= \frac{v}{4}s_{2\beta}\left[-\lambda_1 + \lambda_2 +(\lambda_1+\lambda_2 -2\lambda_{345})c_{2\beta} \right]. 
\end{align}

\end{document}